\documentclass[longauth]{aa}
\usepackage[varg]{txfonts}
\usepackage{natbib}
\bibpunct{(}{)}{;}{a}{}{,} \usepackage{graphicx}	\usepackage{amsmath}	\usepackage{amssymb}	\usepackage{ragged2e}   \usepackage[
    separate-uncertainty=true,
    multi-part-units=single,
    range-phrase=--,
    range-units=single
]{siunitx}
\ifdefined\unit\else
  \ifdefined\NewCommandCopy
    \NewCommandCopy\unit\si
  \else
    \NewDocumentCommand\unit{O{}m}{\si[#1]{#2}}
  \fi
\fi
\usepackage{physics} \usepackage[dvipsnames]{xcolor}
\usepackage{tikz}
\usepackage{pgfplots}
\usetikzlibrary{pgfplots.groupplots}
\usepackage{hyperref}
\usepackage{enumitem}
\usepackage{dblfloatfix}

\DeclareSIUnit{\au}{AU}
\DeclareSIUnit{\dex}{dex}
\DeclareSIUnit{\hjulianday}{HJD}
\DeclareSIUnit{\Jansky}{Jy}
\DeclareSIUnit{\julianday}{JD}
\DeclareSIUnit{\bjd}{BJD}
\DeclareSIUnit{\mas}{mas}
\DeclareSIUnit{\mag}{mag}
\DeclareSIUnit{\parsec}{pc}
\DeclareSIUnit{\permille}{\text{\textperthousand}}
\DeclareSIUnit{\rad}{rad}
\DeclareSIUnit{\solarmass}{M_\odot}
\DeclareSIUnit{\solarradius}{R_\odot}
\DeclareSIUnit{\solarluminosity}{L_\odot}
\DeclareSIUnit{\year}{yr}
\DeclareSIUnit{\earthmass}{M_\oplus}

\bibpunct{(}{)}{;}{a}{}{,}

\defcitealias{gaiaDR3}{GDR3}
\defcitealias{SuarezMascareno2017}{SM17}
\defcitealias{Stefanov2025a}{S25}
\defcitealias{Hara2022}{H22}

\makeatletter
\DeclareRobustCommand
  \shortvdots{\vbox{\baselineskip4\p@ \lineskiplimit\z@
    \hbox{.}\hbox{.}\hbox{.}}}
\makeatother 
\title{
Stellar-activity analysis of the nearby M dwarf GJ~526
}
\subtitle{
Multi-dimensional Gaussian-process modelling of RV, FWHM and S-index
}

\authorrunning{Stefanov et al.} 
\author{
A.~K.~Stefanov\inst{1,2,*},
J.~I.~Gonz\'alez~Hern\'andez\inst{1,2},
A.~Su\'arez~Mascare\~no\inst{1,2},
R.~Rebolo\inst{1,2,3},
N.~Nari\inst{1,2,4},
J.~M.~Mestre\inst{5},
S.~G.~Sousa\inst{6},
H.~M.~Tabernero\inst{7},
M.-R.~Zapatero~Osorio\inst{8},
P.~Figueira\inst{9},
J.~P.~Faria\inst{9},
M.~J.~Hobson\inst{9},
A.~M.~Silva\inst{6,10},
A.~Castro-Gonz\'alez\inst{8},
N.~C.~Santos\inst{6,10},
A.~Sozzetti\inst{11},
F.~Pepe\inst{9},
S.~Cristiani\inst{12,13},
B.~Lavie\inst{9},
C.~J.~A.~P.~Martins\inst{6,14}
}

\institute{
\inst{1}Instituto de Astrof\'isica de Canarias, 38205 La Laguna, Spain\\
\inst{2}Departamento de Astrof\'isica, Universidad de La Laguna, 38206 La Laguna, Spain\\
\inst{3}Consejo Superior de Investigaciones Cient\'ificas (CSIC), 28006 Madrid, Spain\\
\inst{4}Light Bridges S. L., 35004 Las Palmas de Gran Canaria, Spain\\
\inst{5}Dipartimento di Fisica e Astronomia "Galileo Galilei", Universit\`a di Padova, Vicolo dell’Osservatorio 3, 35122 Padova, Italy\\
\inst{6}Instituto de Astrof\'isica e Ci\^encias do Espa\c{c}o, CAUP, Universidade do Porto, Rua das Estrelas, 4150-762 Porto, Portugal\\
\inst{7}Departamento de F\'isica de la Tierra y Astrof\'isica \& IPARCOS-UCM (Instituto de F\'isica de Part\'iculas y del Cosmos de la UCM), Facultad de Ciencias F\'isicas, Universidad Complutense de Madrid, 28040 Madrid, Spain\\
\inst{8}Centro de Astrobiolog\'ia (CAB), CSIC-INTA, ESAC campus, Camino Bajo del Castillo s/n, 28692 Villanueva de la Ca\~nada, Spain\\
\inst{9}Observatoire Astronomique de l’Universit\'e de Gen\`eve, Chemin Pegasi 51b, 1290 Versoix, Switzerland\\
\inst{10}Departamento de F\'isica e Astronomia, Faculdade de Ci\^encias, Universidade do Porto, Rua do Campo Alegre, 4169-007 Porto, Portugal\\
\inst{11}INAF - Osservatorio Astrofisico di Torino, Via Osservatorio 20, 10025 Pino Torinese, Italy\\
\inst{12}INAF - Osservatorio Astrofisico di Trieste, via G. B. Tiepolo 11, 34143 Trieste, Italy\\
\inst{13}IFPU - Institute for Fundamental Physics of the Universe, via Beirut 2, 34151 Trieste, Italy\\
\inst{14}Centro de Astrof\'isica da Universidade do Porto, Rua das Estrelas, 4150-762 Porto, Portugal\\
\inst{*}\email{atanas.stefanov@iac.es}
}

\date{
Received 10 July 2025 / Accepted 2 September 2025
}

\abstract{M dwarfs are the most abundant stars in the Galaxy, and their low masses make them natural favourites for exoplanet radial-velocity (RV) searches. Nevertheless, these stars often demonstrate strong stellar activity that is yet to be fully understood. We use high-precision ESPRESSO, HIRES, and HARPS spectroscopy to perform a stellar-activity analysis on the nearby early M dwarf GJ~526 \mbox{($d=5.4$\,pc)}. We carry out joint modelling of: (i)~stellar rotation in RV, FWHM, and Mount Wilson S-index through Gaussian processes, (ii)~long-term trends in these three physical quantities, (iii)~RV planetary signals. We constrain the stellar-rotation period to \mbox{$P_\text{rot}=48.7\pm 0.3\,$d}, and discover a weak sinusoidal signature in RV, FWHM and S-index of period \mbox{$P_\text{cyc}=1680^{+50}_{-40}$\,d}. We propose phase-space trajectories between RV and activity indicators as a novel means to visualise and interpret stellar activity. Current evidence does not support planetary companions of GJ~526.}

\keywords{
techniques: radial velocities -- stars: activity -- stars: individual: GJ~526
}

\begin{document}
\maketitle

\section{Introduction}

M dwarfs, the most prevalent stars in the Galaxy, remain very attractive for exoplanetary radial-velocity (RV) detection due to their low masses. In practice, however, M dwarfs stubbornly mask potential planetary RV signals through physical processes such as stellar activity on both short and long timescales (\citealp{Borgniet2015} and \citealp{Dumusque2011a} respectively). To compound the situation, M-dwarf activity is sometimes exhibited at periods that overlap with the habitable zone of potential planetary companions (e.g. GJ~15A; \citealp{Pinamonti2018}). It is therefore important to understand better how stellar activity affects RV measurements, so as to improve our ways of detection.

In this work, we examine the RV behaviour of one such M dwarf, GJ~526
\mbox{($\alpha=13\textsuperscript{h}45\textsuperscript{m}43.8\textsuperscript{s}$}, 
\mbox{$\delta=14\degr53'29.5''$}; \citealp{gaiaDR3}, hereafter \citetalias{gaiaDR3}). This nearby star has been known to the scientific community since the beginning of the nineteenth century at the latest \citep{lalande}. Its close proximity to the Sun was identified at least since \citet{Porter1892} measured its extraordinarily large proper motion.
At present, the distance to GJ~526 is well-constrained to \SI{5.435}{\parsec} (\citetalias{gaiaDR3}), but its stellar parameters are moderately established. GJ~526 is regarded to have a spectral class between M1 and M2 \citep{Rojas-Ayala2012,Gaidos2014}; and recently, \citet{Passegger2022} computed standardised stellar parameters and reported an effective temperature $T_\text{eff}=\SI{3648\pm 88}{\kelvin}$ and a surface gravity $\log g= 4.75\pm 0.04$, in combination with past measurements in literature (see references therein). GJ~526 exhibits a particularly slowly evolving and foreseeable activity, which made it an attractive choice for a spectral standard in studies of various nature (e.g. \citealp{Binks2016,Martinez2017,Pancino2017,Fuhrmeister2019}). Its stellar rotation is revealed by periodic signatures in both RV and activity data -- \citet{SuarezMascareno2017}, hereafter \citetalias{SuarezMascareno2017}, identified a strong RV signal at \SI{49.2\pm 0.1}{\day} and constrained the stellar-rotation period to \SI{52.1\pm 12.0}{\day}.

GJ~526 has been studied by several RV programmes, including: The Lick Planet Search Program (hereafter Lick; \citealp{lick}), HIRES \citep{hires}, HARPS \citep{harps}, CARMENES \citep{carmenes} and ESPRESSO \citep{espresso_early,espresso}, in this order. The combination of these surveys provides with \SI{26}{\year} velocimetry, which includes high-precision measurements from the ultra-stable ESPRESSO spectrograph, with a typical photon-noise uncertainty near $\SI{10}{\centi\metre\per\second}$. Despite that all aforementioned sources provide with rich and multifaceted data, there has been no evidence of planetary companions of GJ~526 so far. This work continues as follows. Section~\ref{sec:observations} describes the observational data relevant to our study. Section~\ref{sec:stellar_parameters} contains our derivation of stellar parameters for GJ~526. Section~\ref{sec:modelling} reviews our modelling approach, Section~\ref{sec:analysis} looks into the results of our models, and Section~\ref{sec:discussion} discusses their implications. We summarise our work in Section~\ref{sec:conclusions}.

\section{Observations}\label{sec:observations}
\subsection{ESPRESSO velocimetry}
ESPRESSO is a high-resolution ultra-stable échelle spectrograph. It is located at the ESO's Very Large Telescope, Paranal Observatory, Chile, and it can be fed with signal from either one or all \SI{8.2}{\metre} unit telescopes of the facility \citep{espresso_early,espresso}. This spectrograph was designed to reach an RV precision of \SI{10}{\centi\metre\per\second}, allowing to detect Earth-like planets in the habitable zones of GKM dwarfs. ESPRESSO already achieves an RV precision of \SI{25}{\centi\metre\per\second} over the course of a night and \SI{50}{\centi\metre\per\second} over the course of months. In fact, \citet{espresso} suggests an instrumental precision of \SI{10}{\centi\metre\per\second}, leaving aside photon-noise and stellar-jitter limits.
We acquired 75 ESPRESSO measurements of GJ~526 as part of its Guaranteed Time Observations (GTO).\footnote{
Measurements came from the following programmes:
106.21M2,
108.2254,
110.24CD,
1102.C-0744,
1102.C-0958, and
1104.C-0350.
}
In nine measurements, the atmospheric dispersion corrector of ESPRESSO malfunctioned, which led to a non-quantified error in velocimetry.\footnote{
Those measurements took place near the following \unit\bjd~timestamps:
2459278.8,
2459284.7,
2459289.7,
2459289.9,
2459299.7,
2459313.7,
2459322.7, and
2459327.6.
}
We removed those and continued with the 66 remaining measurements. ESPRESSO RVs were extracted from the spectra through the cross-correlation function method (CCF method; \citealp{Baranne1996}), and through its own Data Reduction Software (DRS), version 3.2.5.

Our data spans from February 2019 to March 2023, with a median interval of \SI{6.0}{\day} between measurements. This temporal coverage includes the fibre link change at June 2019, which introduced an RV offset in measurements \citep{espresso}. This led some works to regard the ESPRESSO dataset as two separate ones: before and after the fibre link (e.g. \citealp{Faria2022,SuarezMascareno2023}). We also follow this procedure, and hereafter label these datasets as ESPRESSO18 \mbox{($N=14$)} and ESPRESSO19 \mbox{($N=52$)}. Both subsets have a similar RV root mean square (rms) and median uncertainty. ESPRESSO18 is characterised by an RV rms of \SI{2.49}{\metre\per\second} and a median uncertainty of \SI{0.13}{\metre\per\second}; while ESPRESSO19 has an RV rms of \SI{2.20}{\metre\per\second} and a median photon-noise uncertainty of \SI{0.14}{\metre\per\second}.

\subsection{HARPS velocimetry}
HARPS is a high-resolution échelle spectrograph located at the \SI{3.6}{\metre} telescope, ESO, La Silla, Chile, and was the first spectrograph  to break the \SI{1}{\metre\per\second} barrier in RV precision \citep{harps}. We made use of the public HARPS RV database by \citet{harps_trifon} that contains 32 HARPS measurements of GJ~526, spanning from May 2005 to June 2009. All of these were acquired before the optical fibre update in May 2015. Two spectroscopic measurements were excluded on account of failing a iterative $3\sigma$-clipping criterion in FWHM. Then, the first and the last point of this intermediate dataset
were separated by \SI{310}{\day} and \SI{1845}{\day} from the bulk of measurements (i.e. 21\% and 125\% relative to the bulk baseline). We chose to mask those out. This procedure yielded 28 spectroscopic measurements that span from May 2005 to June 2009, with a median interval of \SI{3.1}{\day} between measurements. Our HARPS data extend the baseline nicely, and have a median RV uncertainty of \SI{0.62}{\metre\per\second}.

\subsection{HIRES velocimetry}
HIRES is a high-resolution échelle spectrograph located at the \SI{10}{\metre} Keck telescope, Mauna Kea Observatory, HI, USA \citep{hires}, and was initiated in 1994 as a search for exoplanets targeted around F, G, K and M dwarfs.
\citet{hires_survey} contains 56 HIRES measurements of GJ~526 that span from April 1997 to January 2014, with a median interval of \SI{79}{\day} between measurements. We used the systematic-corrected variant of this dataset, provided by \citet{Tal-Or2019}. Their RV data has an rms of \SI{3.43}{\metre\per\second} and a median uncertainty of \SI{1.61}{\metre\per\second}.

\subsection{CARMENES velocimetry}
CARMENES is a high-resolution échelle spectrograph located at the \SI{3.5}{\metre} telescope at the Calar Alto Observatory, Spain, that was specially designed to search for planetary signatures around nearby, cool stars \citep{carmenes_instrument,carmenes}.
There are 253 measurements of GJ~526 in the CARMENES Data Release 1 \citep{Ribas2023}, spanning from January 2016 to April 2018, with a median interval of \SI{1.0}{\day} between measurements, and a median RV uncertainty of \SI{1.51}{\metre\per\second}. We were unable to meaningfully work with this data, and it became necessary to exclude them from analysis. We describe the features of CARMENES data and the issues we encountered with them in Sect.~\ref{sec:carmenes_incompatibility}.

\subsection{Lick velocimetry}
The Lick programme was carried out with the Hamilton Spectrograph and the \SI{3}{\metre} Shane telescope at the Lick Observatory, CA, USA \citep{lick}. The programme ran from 1987 to 2011 and supplied data that was fundamental to our current understanding of exoplanets \citep{Marcy1996,Butler1997,Butler1999}. The Lick programme contains 16 RV measurements of GJ~526 from August~1992 to February~1997, with a median interval of \SI{73}{\day} between them. This dataset is characterised with large RV uncertainties, with a median of \SI{21.21}{\metre\per\second}. While we acknowledge the historical significance of the programme, such uncertainties are incompatible with the purposes of our study. This led us to exclude this dataset from our analysis.

\subsection{Activity indicators}
Active regions perturb the flux- and velocity fields of the stellar disc, thereby changing the shape of observed lines. Consequently, the CCF also changes in shape. Such perturbations are quantified by tracking the evolution of CCF width, depth and symmetry; and their strength can be related to the coverage of the active regions, their contrast and the stellar $v\sin i$. In this work, we use the CCF full width at half maximum (CCF FWHM) as one of our main activity indicators. Two discussed datasets provide with CCF FWHM timeseries: ESPRESSO and HARPS. ESPRESSO18 data is characterised with an FWHM rms of \SI{2.99}{\metre\per\second} and a median uncertainty of \SI{0.26}{\metre\per\second}. For ESPRESSO19, those values stand at \SI{8.15}{\metre\per\second} and \SI{0.29}{\metre\per\second} respectively. HARPS FWHMs come with an rms of \SI{4.49}{\metre\per\second} and a median uncertainty of \SI{1.23}{\metre\per\second}.

The emission intensity of the cores of Ca II H\&K lines is a reliable proxy of the strength of the stellar magnetic field, and thereby, of the stellar-rotation period $P_\text{rot}$ for late-type stars \citep{Noyes1984,Lovis2005}. We use the Mount Wilson \mbox{S-index}
\begin{equation}
    S_\text{MW} = 1.111\times\frac{\Tilde{N}_\text{H}+\Tilde{N}_\text{K}}{\Tilde{N}_\text{V}+\Tilde{N}_\text{R}}+0.0153,
    \label{eq:s_index}
\end{equation}
where $\Tilde{N}_\text{H}$ and $\Tilde{N}_\text{K}$ are triangular passbands centred at \SI{3968.470}{\angstrom} and \SI{3933.664}{\angstrom}, with a common FWHM of \SI{1.09}{\angstrom}; while $\Tilde{N}_\text{V}$ and $\Tilde{N}_\text{R}$ are rectangular passbands centred at \SI{3901.07}{\angstrom} and \SI{4001.07}{\angstrom}, with a common width of \SI{20}{\angstrom} \citep{Vaughan1978}. Three discussed datasets provide with CCF S-index: ESPRESSO, HARPS, and HIRES. ESPRESSO18 is characterised with an S-index rms of \num{40.4e-3} and a median uncertainty of \num{0.14e-3}. For ESPRESSO19, those values stand at \num{125e-3} and \num{0.17e-3} respectively. Our HARPS measurements have an S-index rms of \num{56.9e-3} and a median uncertainty of \num{0.11e-3}, similar to both ESPRESSO18 and ESPRESSO19. Our HIRES S-index measurements have an rms of \num{95.8e-3}, but came with no uncertainties -- and we assumed an uncertainty equal to the median in HARPS (\num{0.11e-3}).

\subsection{ASAS-SN, ASAS, and TESS photometry}
ASAS-SN is an automated photometric programme that looks out for supernovae and other transient events \citep{asassn1,asassn2}. It is comprised of 24 ground-based \SI{14}{\centi\metre} telescopes that are grouped in fours at six sites. ASAS was a program that aimed to perform photometric monitoring over a large sky area \citep{asas}, and its primary goal was to discover and catalogue variable stars of all kinds. TESS is a red-infrared all-sky survey that was specifically designed to detect exoplanetary transits and that continues to deliver precise long-term photometry \citep{tess}. The mission satellite utilises four wide optical cameras, each of which covering a solid angle of
\mbox{$\SI{24}{\deg}\times\SI{24}{\deg}$.}
We give a short self-contained photometry analysis in Appendix~\ref{sec:photometry}. Its key takeaway is that we found no signals we could attribute to stellar activity, except a \SI{1333}{\day} signal that we associated with the magnetic-cycle period $P_\text{cyc}$.

\section{Stellar parameters}\label{sec:stellar_parameters}
We derived the effective temperature $T_\text{eff}$, the surface gravity $\log g$, and the stellar metallicity [Fe/H] through \textsc{SteParSyn}\footnote{\url{https://github.com/hmtabernero/SteParSyn/}} \citep{steparsyn}. We report the following measurements:
\mbox{$T_\text{eff}=3699\pm 16\,\unit\kelvin$},
\mbox{$\log g=4.69\pm 0.05$},
\mbox{$\text{[Fe/H]}=-0.33\pm 0.07\,\text{dex}$}, as well as the total line-broadening velocity
\mbox{$v_\text{broad}=2.26\pm 0.21\,\unit{\kilo\metre\per\second}$}.
For the \textsc{SteParSyn} analysis, we used the line list and model grid described in \citet{Marfil2021}. The stellar radius and mass were derived from the spectroscopic $\log g$ and from the linear mass-radius relation in \citet{Schweitzer2019}. We report:
\mbox{$M=0.505\pm 0.058\,\unit\solarmass$},
\mbox{$R=0.479\pm 0.055\,\unit\solarradius$}.
Table~\ref{tab:stellar_parameters} gives a complete list of stellar parameters.

\begin{table}[t]
\centering
\caption{Stellar parameters of GJ~526.}
\label{tab:stellar_parameters}
\begin{tabular}{lclc}
\hline\hline
\multicolumn{1}{c}{Parameter} &
\multicolumn{1}{c}{Unit} &
\multicolumn{1}{c}{Value} &
\multicolumn{1}{c}{Ref.} \\ \hline
$\alpha$ (J2000) &
-- &
13\textsuperscript{h}45\textsuperscript{m}43.8\textsuperscript{s} &
1 \\
$\delta$ (J2000) &
-- &
+14\degr53'29.5'' &
1 \\
$\mu_\alpha\cos\delta$ &
\unit{\mas\per\year} &
$+1776.006 \pm 0.034$ &
1 \\
$\mu_\delta$ &
\unit{\mas\per\year} &
$-1455.156\pm 0.017$ &
1 \\
$\varpi$ &
\unit{\mas} &
$183.9962 \pm 0.0253$ &
1 \\
$m_\text{V}$ &
\unit{\mag} &
$8.50 \pm 0.05$ &
2 \\
$m_\text{J}$ &
\unit{\mag} &
$5.18 \pm 0.04$ &
3 \\
$m_\text{H}$ &
\unit{\mag} &
$4.775 \pm 0.206$ &
3 \\
$m_\text{K}$ &
\unit{\mag} &
$4.415 \pm 0.017$ &
3 \\
$T_\text{eff}$ &
\unit{\kelvin} &
$3699\pm 16$ &
0 \\
$\log g$ (cgs) &
-- &
$4.69\pm 0.05$ &
0 \\
$M_\star$ &
\unit{\solarmass} &
$0.505\pm 0.058$ &
0 \\
$R_\star$ &
\unit{\solarradius} &
$0.479\pm 0.055$ &
0 \\
\text{[Fe/H]} &
\unit{\dex} &
$-0.33\pm 0.07$ &
0 \\
$\log R'_\text{HK}$ &
-- &
$-5.286\pm 0.087$ &
0 \\
$P_\text{rot}$ &
\unit{\day} &
$48.7\pm 0.3$ &
0 \\
$v_\text{broad}$ &
\unit{\kilo\metre\per\second} &
$2.26\pm 0.21$ &
0 \\
\hline
\end{tabular}
\tablefoot{
The stellar-rotation period $P_\text{rot}$ comes from our Gaussian-process modelling, and not directly from raw spectra.
}
\tablebib{
(0) This work;
(1) \citet{gaiaDR3};
(2) \citet{VanBelle2009};
(3) \citet{2mass}.
}
\end{table} 
We cross-checked our $T_\text{eff}$, $\log g$ and [Fe/H] results through \textsc{ODUSSEAS}\footnote{\url{https://github.com/AlexandrosAntoniadis/ODUSSEAS}} \citep{odusseas} on the combined high-resolution ESPRESSO spectra. The trigonometric surface gravity for this star was derived using recent \citetalias{gaiaDR3} data and following \citet{Sousa2021}. This derivation yielded:
\mbox{$T_\text{eff}=3656\pm 92\,\unit\kelvin$},
\mbox{$\log g=4.67\pm 0.09$},
\mbox{$\text{[Fe/H]}=-0.242\pm 0.111\,\text{dex}$}.
These measurements are all consistent with our primary stellar-parameter derivation.

We validated our derivations of $M$, $R$, and the bolometric stellar luminosity $L$ by computing the photometric spectral energy distribution (SED) of GJ~526 using the publicly available broad-band photometry from the following catalogues:
GALEX \citep{galex},
TYCHO \citep{tycho2},
Pan-STARRS \citep{panstarrs1},
SDSS \citep{sdss},
\citetalias{gaiaDR3},
2MASS \citep{2mass},
WISE \citep{wise},
IRAS \citep{iras}, and
AKARI \citep{akari}.
We used the photometric zero points listed in the Virtual Observatory SED Analyzer tool (VOSA; \citealp{Bayo2008}) to convert the observed magnitudes into fluxes, and the \citetalias{gaiaDR3} trigonometric parallax to transform from observed to absolute fluxes. Figure~\ref{fig:mariarosa} compares the aforementioned photometric data against a PHOENIX model for solar metallicity and \SI{3700}{\kelvin} \citep{Allard2012}. This fit agrees with data, with the exception of the GALEX datapoint, that is, the bluest photometric point. This goes on to suggest that GJ~526 is an active star.
We then integrated the observed SED, excluding the GALEX datapoint, and obtained
\mbox{$L=0.0401 \pm 0.0027\,\unit\solarluminosity$.}
Together with our derived $T_\text{eff}=3699\pm 16\,\unit\kelvin$ from \textsc{SteParSyn}, we derived
\mbox{$R=0.487\pm 0.029\,\unit\solarradius$} through the Stefan-Boltzmann law. The mass-radius relation of \citet{Schweitzer2019} suggests a stellar mass of
\mbox{$M=0.490\pm 0.030\,\unit\solarmass$.}
The computed surface gravity from these mass- and radius values
\mbox{(log\,$g$ = $4.753 \pm 0.024 $)} agree with our own spectroscopic derivation
\mbox{(log\,$g$ = $4.69 \pm 0.09$)}. Finally, the mass -- $K$-band luminosity relationship of \citep{Mann2019} returned
\mbox{$M=0.483\pm 0.026\,\unit\solarmass$} for GJ~526, again in good agreement with our derivation.

ESPRESSO spectra delivered $\log_{10}R'_\text{HK}$ with a median value of \num{-5.286}, an rms of \num{0.087}, and a median instrumental uncertainty of \num{1.2e-4}. We took the mean and the rms, and reported it in Table~\ref{tab:stellar_parameters}. Through the relation in \citet{SuarezMascareno2018}, our $\log_{10}R'_\text{HK}$ would translate to a stellar-rotation period of
\mbox{$P_\text{rot}=51^{+76}_{-30}\,\unit\day$}, a value that was consistent with our results to follow
\mbox{($P_\text{rot}=48.7\pm 0.3\,\unit\day$}; Sect.~\ref{sec:discussion_activity})

\section{Modelling and inference}\label{sec:modelling}
We use the modelling framework described in \citet{Stefanov2025a}, hereafter \citetalias{Stefanov2025a}, that is generalised for $N$ data sources (e.g. HIRES, HARPS; \mbox{$i\in[0..N-1]$}) and $M$ physical quantities (e.g. RV, FWHM; \mbox{$j\in[0..M-1]$}). The \citetalias{Stefanov2025a} framework includes:
(i) long-term functions (LTFs) for data that follow physical or instrumental trends,
(ii) offsets between different sources,
(iii) source-dependent jitters for each quantity,
(iv) Keplerian- and circular-orbit fitting,
(v) stellar activity modelling using Gaussian processes. A brief discussion of the last step follows.

Gaussian processes (GPs) are collections of random variables, any finite number of which follow a joint Gaussian distribution \citep{Rasmussen2006}. GPs are non-parametric models, meaning that datasets themselves determine the functional form of the model. Instead, we feed to the GP certain functional relationships, which are assumed to describe the correlations between individual measurements. For this reason, GPs are useful devices in the toolbox of planet-detection researchers, as they require no other assumptions of the active regions that the star may exhibit \citep{Haywood2015}.

Stellar activity is often modelled under the assumption that correlations between measurements are sensitive to a certain period -- the stellar-rotation period -- and at the same time, that correlations diminish exponentially at a certain timescale. These two assumptions motivate the squared-exponential periodic (SEP) kernel,\footnote{Also known as the quasi-periodic (QP) kernel in literature.} which has the form
\begin{equation}
    k_\text{SEP}(\Delta t) = \kappa^2\exp\left[
    -\frac{\Delta t^2}{2\tau^2}
    -\frac{\sin^2\left(\pi\Delta t/P\right)}{2\eta^2}
    \right],
    \label{eq:sep_kernel}
\end{equation}
where $\kappa^2$ is the maximum amplitude of the kernel, $\tau$ is the timescale of coherence, $P$ is associated with the stellar-rotation period $P_\text{rot}$, and $\eta$ describes the kernel-feature complexity \citep{Haywood2014,Rajpaul2015,Angus2018}. We refer to $\eta$ as the `sinescale' on account of its algebraic analogy to the timescale $\tau$. Invoking the SEP kernel directly comes with a computational overhead, and we work with two of its approximations: the Matérn 3/2 exponential periodic (MEP) kernel and the exponential-sine periodic (ESP) kernel, both as introduced in the \textsc{s+leaf}\footnote{\url{https://gitlab.unige.ch/delisle/spleaf}} library \citep{spleaf1, spleaf2}.

Real-data experience shows that conditioning GP kernels onto RV data alone may cause the model to incorrectly address planetary signals as stellar activity, and therefore hinder planet detections. An emerging way to combat this is to fit GP-based models on RV- and activity-indicator data, i.e. in several dimensions at the same time. By `informing' the GP kernel of the activity-indicator behaviour of the star, we allow that kernel to better separate planetary from non-planetary signals in RV \citep{Ahrer2021,Rajpaul2021}. In the case of GJ~526, we worked with RV, FWHM and S-index at the same time. We followed the notation of \citetalias{Stefanov2025a}, and assigned \mbox{$j=0$} for RV, \mbox{$j=1$} for FWHM, and \mbox{$j=2$} for S-index.

There are different ways to assign GPs to several physical quantities. One of them, the multi-dimensional regime, follows the $FF'$ formalism that was defined in \citet{Aigrain2012} and extended in \citet{Rajpaul2015}. In this regime, one GP kernel is fitted on all physical quantities at the same time. For measurements $y_{i,j}$ at times $t_{i,j}$, we solve the system of equations
\begin{equation}
    \left|\begin{array}{l}
    y_{i,j}=A_j\,G(t_{i,j})+B_j\,(\mathrm{d}G/\mathrm{d}t)_{t=t_{i,j}}\\[0.2ex]
    \shortvdots
    \end{array}\right.\qquad
    \begin{array}{l}
        \forall\, i, \\
        \forall j,
    \end{array}
    \label{eq:multidimensional_system}
\end{equation}
where $A_j$ and $B_j$ are quantity-specific fit hyperparameters, and $G(t_{i,j})$ is the kernel estimation of the stellar activity at given times. The multi-dimensional regime is a common approach for the disentanglement of stellar activity in the velocimetry of late-type dwarfs \citep{Barragan2023,Passegger2024}, and it is useful to break degeneracies in cases where there is a significant correlation between physical quantities \citep{SuarezMascareno2020}. \citet{Rajpaul2015} discussed that the $FF'$ formalism expects only some physical quantities to have non-zero $B_j$ terms in \eqref{eq:multidimensional_system}; we refer to those as `gradiented' with respect to the remaining quantities. We use the \textsc{s+leaf} MultiSeriesKernel procedure to realise this regime.

\subsection{Parameter priors}\label{sec:parameter_priors}
We use a set of default parameter priors across models unless stated explicitly. Our default LTF parameter priors are informed by the solutions of the Levenberg-Marquadt algorithm (LMA; \citealp{Levenberg1944,Marquardt1963}). We run an instance of the LMA that returns a mean $\mu_\text{lm}$ and an uncertainty $\sigma_\text{lm}$ for each LTF parameter. For these parameters, we then use the priors $\mathcal{N}\left(\mu_\text{lm},200\sigma_\text{lm}\right)$. The large scale factor of $\sigma_\text{lm}$ accounts for the minute errors that the LMA tends to return.

Our default offset priors are based on the common standard deviation of all $y_{i,j}$, which we denote $\sigma_j$. The default prior on offsets is \mbox{$O_{i,j}=\mathcal{N}(0,5\sigma_j)$}. For jitters, we use the non-restrictive \mbox{$J_{i,j}=\mathcal{U}_\text{log}(10^{-3},10^{3})$}, except for RV jitters, where we take the log-normal distribution
\mbox{$O_{i,0}=\exp\left[\mathcal{N}(\ln\sigma_0,\ln\sigma_0)\right]$}, following \citet{GonzalezHernandez2024}. In terms of SEP kernel hyperparameters, we use timescale priors of
\mbox{$\mathcal{U}_\text{log}(20,10^{4})\,\unit\day$},
period priors of
\mbox{$\mathcal{U}(40.1,64.1)\,\unit\day$} as informed by \citetalias{SuarezMascareno2017} 
\mbox{($52.1\pm 12.0\,\unit\day$)},
sinescale priors of
\mbox{$\mathcal{U}_\text{log}(10^{-2},10^{2})$},
and the following amplitudes:
\mbox{$A_0,B_0= \mathcal{U}(-10^{-3},10^{3})$} for GP RV amplitudes, and
\mbox{$A_j= \mathcal{U}_\text{log}(10^{-3},10^{3})$} for GP indicator amplitudes.

\subsection{Inference}

\citet{Skilling2004} noted that while directly sampling from the likelihood function $\mathcal{L}$ is exponentially more expensive, sampling uniformly within a bound \mbox{$\mathcal{L}>k$} is much easier -- and increasing $k$ iteratively upon reaching convergence can be used to evaluate the posterior. This strategy is now known as nested sampling. Its formalism allows it to have:
(i) lower computational complexity;
(ii) the ability to explore multi-modal distributions of arbitrary complex shapes;
(iii) the ability to numerically evaluate the Bayesian evidence $Z$. We used \mbox{ReactiveNestedSampler}, a nested-sampling integrator provided by \textsc{UltraNest}\footnote{\url{https://github.com/JohannesBuchner/UltraNest}} \citep{ultranest}. In every inference instance of $N_\text{param}$ model parameters, we required $40N_\text{param}$ live points and a region-respecting slice sampler that accepted $4N_\text{param}$ steps until the sample was considered independent.

\section{Analysis}\label{sec:analysis}
\subsection{Features of available velocimetry}
\begin{figure*}
    \centering
    \resizebox{\hsize}{!}{\includegraphics{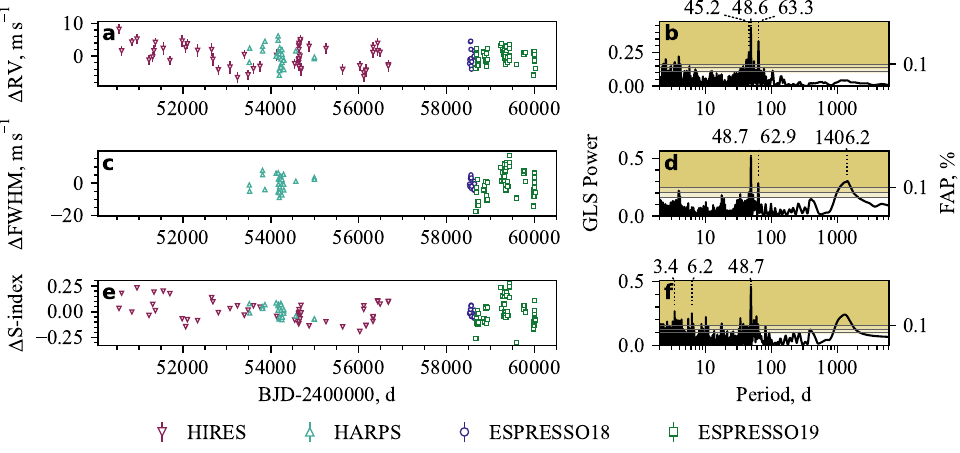}}
    \caption{
    Timeseries of mean-subtracted:
    (a) RV,
    (c) FWHM,
    (e) S-index. Measurements are marked depending on the data source: pink downward triangles for HIRES, teal upward triangles for HARPS, blue circles for ESPRESSO18, and green squares for ESPRESSO19. (b,d,f) Associated wide-period GLSPs of RV and activity indicators. Three FAP levels: 10\%, 1\% and 0.1\%, split GLSP ordinates in bands of different colour. We highlight the three most prominent peaks in each GLSP.
    }
    \label{fig:timeseries}
\end{figure*}

Figure \ref{fig:timeseries} shows the generalised Lomb-Scargle periodograms (GLSPs; \citealp{Lomb1976,Scargle1982,Zechmeister2009,VanderPlas2015}) of RV and all activity indicators in the period range \SIrange{2}{6000}{\day}. RV data seem to be devoid of long-term trends. The GLSPs of FWHM and S-index, on the other hand, reveal significant periodicities near \SI{1400}{\day} in both physical quantities (Figs.~\ref{fig:timeseries}d,f). This may be indicative of a magnetic cycle, and as such, it calls to be potentially included in the LTFs of our chosen models.

At shorter periods, we observe a distinct \SIrange{48}{49}{\day} signal in RV and both activity indicators. This matches the stellar-rotation period measured by \citetalias{SuarezMascareno2017} ($P_\text{rot}=52.1\pm 12.0\,\unit\day$), and we likewise associate it to $P_\text{rot}$. There are other signals in its company, most notably: (i) a \SIrange{3}{4}{\day} signature in RV, FWHM and S-index, (ii) a forest of peaks extending up to \SI{10}{\day} in S-index, (iii) a \SI{63}{\day} signature in RV and FWHM. A priori, we would expect that these may be aliases of $P_\text{rot}$, which come from the harmonic combination of $P_\text{rot}$ and any significant window-function (WF) peak at $P_\text{wf}$ through the relation
\begin{equation}
    P_\text{alias}^{-1}=\left|P_\text{rot}^{-1}\pm P_\text{wf}^{-1}
    \right|.
    \label{eq:alias_equation}
\end{equation}
as described in \citet{Dawson2010}. Therefore, by identifying the dominant WF peaks, we can check for aliases of $P_\text{rot}$.

HIRES supplies with RV and S-index data, but not with FWHM. Consequently, the WF formed by FWHM measurements is not equivalent to the WF that RV and S-index time series share. Although the timestamps of HIRES measurements certainly relocate powers in the WF, we verified that the first few WF peaks change marginally with the inclusion of HIRES points. We continued with the WF formed by HIRES, HARPS and ESPRESSO to select alias-inducing peaks. Figure~\ref{fig:alias_analysis} visualises our alias analysis. We identified four definite WF peaks that inject aliases of $P_\text{rot}$. Those WF peaks are located at: \SI{215}{\day}, \SI{365}{\day}, \SI{477}{\day} and \SI{657}{\day} (Fig.~\ref{fig:alias_analysis}g). For $P_\text{rot}$, they would generate aliases between \SIrange{30}{70}{\day}. We observed that these WF peaks indeed generate aliases that match observed structures, including the 0.1\% FAP \SI{63}{\day} in RV, FWHM and S-index (Figs.~\ref{fig:alias_analysis}a,b,c, teal upward triangle). A small peak between \SIrange{30}{40}{\day} is also identified as an alias in all dimensions. The forest of signatures between \SIrange{1}{10}{\day} remains unidentified in all physical quantities.

\subsection{Long-term function selection}\label{sec:ltf_selection}
What remains to be verified is whether the \SI{1400}{\day} signals proposed by both FWHM and S-index GLSPs are statistically significant. This can be checked through a suite of intermediate models that fit different LTFs to one-dimensional time series, without including any GPs. We prepared models that incorporate a polynomial function of degree \mbox{$p_j\leq 2$} with respect to time, as well as a potential sinusoidal term. The full form of this general LTF is
\begin{equation}
    f_j(t_{i,j}) =
    \sum_{n=0}^{p_j}
    \alpha_n t^n_{i,j} +
    k_{\text{cyc, }j} \sin\left[\
    \frac{2\pi(t_{i,j}+P_\text{cyc}\,\varphi_{\text{cyc, }j})}{P_\text{cyc}}
    \right],
    \label{eq:ltf_equation}
\end{equation}
where $f_j$ is the LTF as defined in \citetalias{Stefanov2025a}, $k_{\text{cyc, }j}$ and $\varphi_{\text{cyc, }j}$ are the cycle amplitude and cycle phase, and $P_\text{cyc}$ is the cycle period.\footnote{This implies the two conditions assumed in \citetalias{Stefanov2025a}: \mbox{(i) $\forall i,j:\operatorname{avg} y_{i,j}=0$}, \mbox{(ii) $\max{t_{i,j}}=0$}; see Sect.~3 of the same work for more detail.}
We used a cycle-period prior of \mbox{$\mathcal{U}_\text{log}(200,10^4)\,\unit\day$}. For models with a sine-free LTF, we set \mbox{$k_{\text{cyc, }j}=0$} and excluded the parameters of this term from sampling.

Zero-order polynomial LTFs scored best among all dimensions in terms of Bayesian evidence. This matches the lack of signals beyond \SI{4000}{\day} in GLSPs (Fig.~\ref{fig:timeseries}b,d,f). The inclusion of a sine term in the LTF is favoured in FWHM
\mbox{($\Delta\ln Z = 2.8$)}
as well as S-index
\mbox{($\Delta\ln Z = 4.7$)}.
These intermediate models successfully recover a
\mbox{$1380^{+50}_{-290}\,\unit\day$} FWHM signal of amplitude
\mbox{$4.59^{+1.13}_{-1.27}\,\unit{\metre\per\second}$}, and a
\mbox{$1350^{+1400}_{-30}\,\unit\day$} S-index signal of amplitude
\mbox{$(5.77^{+1.19}_{-1.21})\times\num{e-2}$}. The phase posteriors from these two models hint that the two signals may be in phase ($0.70^{+0.17}_{-0.08}$ in FWHM, $0.73^{+0.07}_{-0.20}$ in S-index). The agreement of these solutions with the \SI{1400}{\day} peaks in the GLSPs and the $\Delta\ln Z$ improvement from their non-sine counterparts steers us towards a global zero-order LTF with a sine term.

\subsection{Model selection}
Having gained insights from Sect.~\ref{sec:ltf_selection}, we adopted the LTF
\begin{equation}
    f_j(t_{i,j}) =
    \alpha_i +
    k_{\text{cyc, }j} \sin\left[\
    \frac{2\pi(t_{i,j}+P_\text{cyc}\,\varphi_{\text{cyc, }j})}{P_\text{cyc}}
    \right].
    \label{eq:model_ltf}
\end{equation}
We used the following LTF priors: a sine-term period of \mbox{$P_\text{cyc}=\mathcal{U}(1000,2000)\,\unit\day$}, and the following semi-amplitudes:
\mbox{$k_\text{cyc,0}=\mathcal{U}(0,10)\,\unit{\metre\per\second}$} in RV,
\mbox{$k_\text{cyc,1}=\mathcal{U}(0,10)\,\unit{\metre\per\second}$} in FWHM, and
\mbox{$k_\text{cyc,2}=\mathcal{U}(0,0.1)$} in S-index. Then, we built a grid of multi-dimensional GP models that fit in all three physical quantities. This model grid probed different stellar-activity kernels, as well as different planetary configurations. For stellar activity, we used four different approximations of the SEP kernel: MEP, as well as ESP with 2, 3 and 4 harmonics (hereafter ESP2, ESP3 and ESP4). The rank of these kernels increase in this order, and they can be assumed to grow in complexity likewise. In the design of planetary configurations, we either modelled for no planets, or for a single circular-orbit planet. This gave a grid of eight different models. In all models, we used the multi-dimensional regime, with only RV being gradiented relative to FWHM and S-index. We used the default parameter priors as described in Sect.~\ref{sec:parameter_priors}. For the modelled planet, we adopted an uninformed orbital-period prior of \mbox{$P_\text{b}=\mathcal{U}_\text{log}(1,1000)\,\unit\day$} and an amplitude prior of
\mbox{$k_\text{rv, b}=\mathcal{U}(0,5)\,\unit{\metre\per\second}$}.

\begin{figure}
    \centering
    \resizebox{0.95\hsize}{!}{\includegraphics{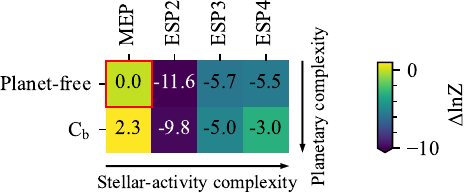}}
    \caption{
    Bayesian-evidence comparison between different planetary configurations and stellar activity kernels. Planetary configurations include a circular-orbit component (C$_\text{b}$; $e=0$). For stellar activity, we utilised the MEP kernel and the ESP kernel with 2, 3 and 4 harmonics. The model that we further elect for analysis assumed a planet-free model with a MEP kernel (red border; $\ln Z=-434.4$).
    We give the Bayesian factor $\Delta\ln Z$ of remaining models relative to this model.
    }
    \label{fig:model_grid}
\end{figure}

Figure~\ref{fig:model_grid} displays a $\ln Z$ comparison of entries throughout our model grid. Entries are arranged so that planetary-system complexity increases downwards, and stellar-activity increases rightwards. The MEP kernel is widely endorsed by the model grid, with an average improvement in evidence of
\mbox{$(\Delta\ln Z)_\text{avg}>5.4$} 
relative to ESP kernels. The inclusion of a single circular-orbit planet is not favoured much by data -- we report an average improvement in evidence of 
\mbox{$(\Delta\ln Z)_\text{avg}=1.8$} across all kernels, with the largest improvement being
\mbox{$\Delta\ln Z=2.5$} for the ESP4 kernel. We thereby select the most appropriate (`best') model in our analysis: the MEP planet-free model. What now follows is a discussion of the best-model fit and a brief comparison with similar models in our grid. We cover the implications of the best-model results in Sect.~\ref{sec:discussion_activity}.

\begin{figure}
    \centering
    \resizebox{\hsize}{!}{\includegraphics{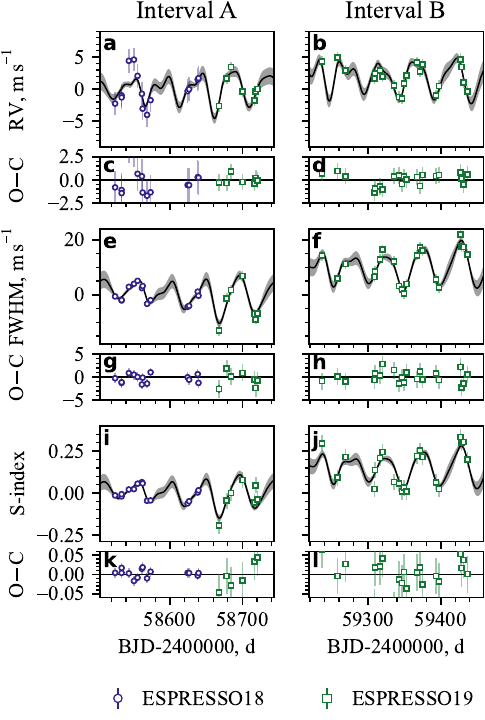}}
    \caption{
    Raw time series (markers) against our best model (black; LTF+GP) in two selected intervals. Data points come with two error bars: one assigned by the instrument (saturated), and another that includes the model jitter (semi-transparent). We provide with a direct comparison and associated O-C diagrams in:
    (a,b,c,d) RV,
    (e,f,g,h) FWHM,
    (i,j,k,l) S-index.
    }
    \label{fig:bestmodel_fit}
\end{figure}

Our combined dataset of HIRES, HARPS, and ESPRESSO offers a very dissimilar temporal sampling. There are clusters of dozens of points within a few hundred days, as well as measurement gaps over several years. Although this uneven sampling is expected to challenge the GP framework, we observe a sound prediction in all physical quantities in the full baseline, as well as in clusters of densely sampled points. Figure~\ref{fig:bestmodel_fit_full} displays the prediction of our best model (an LTF and a GP) against all raw time series. We observe stable structures, and no significant residual signals in any dimension. Figure~\ref{fig:bestmodel_fit} displays the same prediction, but in two selected densely sampled \SI{240}{\day} intervals: Interval A (approx. \SIrange{2458504}{2458744}{\bjd}), and Interval B (approx. \SIrange{2459219}{2459459}{\bjd}), both with \SI{0.1}{\day} prediction sampling. The MEP kernel models all measurements reasonably well, sans a few RV points in ESPRESSO18 (Fig.~\ref{fig:bestmodel_fit}a). All \mbox{O-C} diagrams reveal residuals that are well distributed around the zero, with a spread similar to the assigned model jitters. We report the following residual rms against the median `jittered' uncertainties:
\SI{1.46}{\metre\per\second} against \SI{1.89}{\metre\per\second} in RV,
\SI{2.57}{\metre\per\second} against \SI{1.85}{\metre\per\second} in FWHM, and
\num{29e-3} against \num{33e-3} in S-index. We supply the posterior of our best model in Table~\ref{tab:bestmodel_posterior}, under column `All-data'. All posteriors were well-behaved in shape and did not appear to be biased by the prior space we imposed. We discuss the meaning of these results in Sect.~\ref{sec:discussion_activity}.

We compared our best model with other planet-free models in our model grid (Fig.~\ref{fig:model_grid}), that is, against ESP2, ESP3 and ESP4 models. Table~\ref{tab:bestmodel_posterior_comparison} lists the posteriors of the four aforementioned models. We report marginal differences between all posteriors overall, with the exception of a larger timescale in MEP
\mbox{($\tau=277^{+109}_{-82}\,\unit\day$)} than in ESP kernels:
\mbox{$68^{+22}_{-19}\,\unit\day$} in ESP2,
\mbox{$79^{+23}_{-20}\,\unit\day$} in ESP3, and
\mbox{$81^{+22}_{-18}\,\unit\day$} in ESP4. This difference might come from how $\tau$ enters algebraically into the MEP and the ESP kernels (see \citealp{spleaf2}). 
While our best model assigned a larger $\tau$ uncertainty compared to its ESP counterparts, it constrained the stellar-rotation period $P_\text{rot}$ better. We report
\mbox{$P_\text{rot}=48.7\pm 0.3\,\unit\day$} in MEP, against:
\mbox{$49.2^{+1.7}_{-1.5}\,\unit\day$} in ESP2,
\mbox{$49.2^{+1.2}_{-0.9}\,\unit\day$} in ESP3, and
\mbox{$49.3^{+1.1}_{-0.9}\,\unit\day$} in ESP4.

Then, we took our best model, and reduced it to two two-dimensional partitions: RV\,\&\,FWHM, and RV\,\&\,S-index. Albeit with marginally noisier posteriors, these two partitions remained potent in the modelling of stellar activity. We report the following stellar-rotation period and timescale for RV\,\&\,FWHM and RV\,\&\,S-index respectively:
\mbox{$48.8^{+0.3}_{-0.4}\,\unit\day$} and \mbox{$48.7^{+0.5}_{-0.6}\,\unit\day$} for $P_\text{rot}$, as well as
\mbox{$353^{+489}_{-185}\,\unit\day$} and \mbox{$147^{+78}_{-47}\,\unit\day$} for $\tau$. Other partition-model parameters have a similar posterior to those from our best model, with the exception of the LTF period, where partition models assigned up to two secondary modes in the \mbox{$\mathcal{U}(1000,2000)\,\unit\day$} prior.

Finally, we used \textsc{s-BART}\footnote{\url{https://github.com/iastro-pt/sBART}} \citep{sbart} on ESPRESSO and HARPS data, so as to self-validate our results under a different reduction framework. We reduced all available raw ESPRESSO and HARPS spectra through \textsc{s-BART}, and from there, we took only RVs and its associated uncertainties. Then, we crossmatched our dataset with \textsc{s-BART} RVs by time, thereby rejecting the same outliers. We report the following rms and median uncertainty of \textsc{s-BART} RVs:
\SI{2.37}{\metre\per\second} and \SI{0.09}{\metre\per\second} for ESPRESSO18,
\SI{2.35}{\metre\per\second} and \SI{0.10}{\metre\per\second} for ESPRESSO19, and
\SI{2.41}{\metre\per\second} and \SI{0.51}{\metre\per\second} for HARPS.
We ran a duplicate of our best model, though with \textsc{s-BART} RVs. We found no differences from our best-model results (Table~\ref{tab:bestmodel_posterior}) that warrant discussion.

\subsubsection{ESPRESSO measurements}

We tested whether ESPRESSO data alone are sufficient to characterise the overall stellar activity. Figure~\ref{fig:esprmodel_fit_full} displays the prediction of our best model that was re-run only on ESPRESSO data. No significant signals remain in the residuals except two FWHM signatures at \SI{1.4}{\day}, \SI{1.8}{\day} and \SI{3.4}{\day} (Fig.~\ref{fig:esprmodel_fit_full}d). These three signals were assigned a false-alarm probability (FAP; \citealp{Baluev2008}) smaller than 1\%.

We supply with the posterior of this variant of the best model in Table~\ref{tab:bestmodel_posterior}, under column `ESPRESSO'. We observed only two appreciable differences between the ESPRESSO-only and the full-data best model. Firstly, the LTF cycle was not well constrained in the ESPRESSO-only model, with a
\mbox{$P_\text{cyc}=1720^{+190}_{-210}\,\unit\day$} and a posterior shape that was sharply cut off by the prior
\mbox{$\mathcal{U}(1000,2000)\,\unit\day$}. This is likely because the \SI{1500}{\day} baseline of ESPRESSO data falls short of the inferred cycle period (\mbox{$1670^{+50}_{-40}\,\unit\day$}). The remaining LTF parameters deteriorated likewise. Secondly, the ESPRESSO-only model inferred a larger sinescale of
\mbox{$\eta=0.94^{+0.78}_{-0.32}$} (against \mbox{$0.53^{+0.13}_{-0.10}$}; Table~\ref{tab:bestmodel_posterior}), which is, nevertheless, an $1\sigma$ agreement. We conclude that inference on just these 65 ESPRESSO measurements returns very similar results to our all-data fits, except when it comes to LTF parameters. This affirms the central role of ESPRESSO data in the analysis.

\subsubsection{CARMENES measurements}\label{sec:carmenes_incompatibility}

We tried to add CARMENES RV and FWHM measurements to our composit dataset, and to run the model grid anew. We could find two apparent differences between CARMENES and the rest of the data: larger RV and FWHM scatter compared to other time series, and strong unexplained signals in both RV and FWHM (Fig.~\ref{fig:carm_timeseries}). Despite having the largest median FWHM uncertainties (\SI{2.01}{\metre\per\second}) and the second largest median RV uncertainties (\SI{1.51}{\metre\per\second}), the inclusion of CARMENES data in either dimension (or both) gave noticeably smaller timescales -- and in most cases, posteriors converged to the lower end of the
\mbox{$\mathcal{U}_{\log}\left(20,10^4\right)\,\unit\day$} prior. Inferring our best model on CARMENES data alone suffered from similar problems, and in this instance even the stellar-rotation period was not identified in its wide \mbox{$\mathcal{U}(40.1,64.1)\,\unit\day$} prior (Fig.~\ref{fig:carmmodel_fit_full}). We attempted to pre-process CARMENES data in the different ways before inference, including:
(i) iterative $3\sigma$ clipping in RV and FWHM,
(ii) nightly binning measurements,
(iii) RVs with/without nightly-zero-point corrections \citep{Ribas2023},
(iv) manual derivation of CCF FWHMs. None of this helped solving the issue at hand, and this led to the exclusion of CARMENES data in our analysis.

\section{Discussion}\label{sec:discussion}
\subsection{Stellar activity of GJ~526}\label{sec:discussion_activity}
The stellar activity of GJ~526 in FWHM and S-index demonstrates a strong positive correlation (\mbox{$R=0.85$}; \citealp{Pearson1895}), which is indeed apparent in Fig.~\ref{fig:timeseries}. As such, FWHM and S-index fit nicely within the $FF'$ formalism, where they are modelled with the same kernel estimation $G(t_{i,j})$ but under different scaling. Through our best model, we report a timescale of
\mbox{$\tau=277^{+109}_{-82}\,\unit\day$} and a stellar-rotation period of
\mbox{$P_\text{rot}=48.7\pm 0.3\,\unit\day$}. This implies the ratio
\mbox{$\tau/P_\text{rot}=5.7^{+2.2}_{-1.7}\,\unit\day$}, which hints towards a light-curve morphology that is not Sun-like (i.e. \mbox{$\tau/P_\text{rot}\nsim 1$}; \citealp{Giles2017}). However, albeit disfavoured by the evidence, ESP models inferred very different timescales that suggest a Sun-like morphology: we report \mbox{$\tau/P_\text{rot}$} values of 
\mbox{$1.4\pm 0.4$} for ESP2,
\mbox{$1.6^{+0.5}_{-0.4}$} for ESP3, and
\mbox{$1.6\pm 0.4$} for ESP4.

We report a sinescale of
\mbox{$\eta=0.53^{+0.13}_{-0.10}$}, and this measurement appears stable for all planet-free models. We observe positive $A_0$, $B_0$, $A_1$, $A_2$ for all models. This hints that stellar-activity modulations correlate positively between RV and activity indicators, as well as between RV and the derivatives of the latter. Furthermore, \mbox{$A_0\ll B_0$}, meaning that stellar activity is more prominent in the RV gradient. These results can be interpreted in the following way. FWHM correlates positively with S-index, meaning that the former becomes wider when there are more active regions facing the observer. On the other hand, FWHM correlates negatively with the stellar flux \citep{SuarezMascareno2020}. This suggests that the active regions themselves are dark spots on the stellar surface. The positiveness of $B_0$ goes on to imply that the RV gradient is also correlated positively with FWHM (and S-index), which would also explain this activity as a dark-spots flux effect.

We detect a weak cycle of period
\mbox{$1680^{+50}_{-40}\,\unit\day$} with the following semi-amplitudes:
\mbox{$0.51^{+0.35}_{-0.33}\,\unit{\metre\per\second}$} in RV,
\mbox{$2.73^{+1.51}_{-1.32}\,\unit{\metre\per\second}$} in FWHM, and
\mbox{$0.019^{+0.019}_{-0.013}$} in S-index. Our cycle-phase uncertainties make it difficult to determine whether cycle components are in, or out of phase -- we report the following absolute phase differences:
\mbox{$0.65^{+0.24}_{-0.18}$} between RV and FWHM,
\mbox{$0.83^{+0.22}_{-0.32}$} between RV and S-index, and
\mbox{$0.08^{+0.24}_{-0.14}$} between FWHM and S-index.
What we measure for $P_\text{cyc}$ is approximately half of \citet{SuarezMascareno2016a}
(\mbox{$9.9\pm 2.8\,\unit\year$}) that had been done from 281 photometric measurements over \SI{6.5}{\year}. Recently, \citet{IbanezBustos2025} measured a similar value (\mbox{$3739\pm 288\,\unit\day$}) from 47 S-index measurements over \SI{12}{\year}. Our FWHM \mbox{($N=91$)} and S-index data \mbox{($N=147$)} over almost \SI{26}{\year} show preference towards a shorter cycle, both at the GLSP stage (Fig.~\ref{fig:timeseries}) as well as during our one-dimensional LTF selection (Sect.~\ref{sec:ltf_selection}). Desaturated and de-mooned ASAS-SN photometry reveals a \SI{1333}{\day} signature that would agree with our measurement of $P_\text{cyc}$ (Sect.~\ref{sec:photometry_asassn}).

\subsection{Phase-space trajectories in RV, FWHM, and S-index}\label{sec:rv_activity_trajectories}

\begin{figure*}
    \centering
    \resizebox{\hsize}{!}{\includegraphics{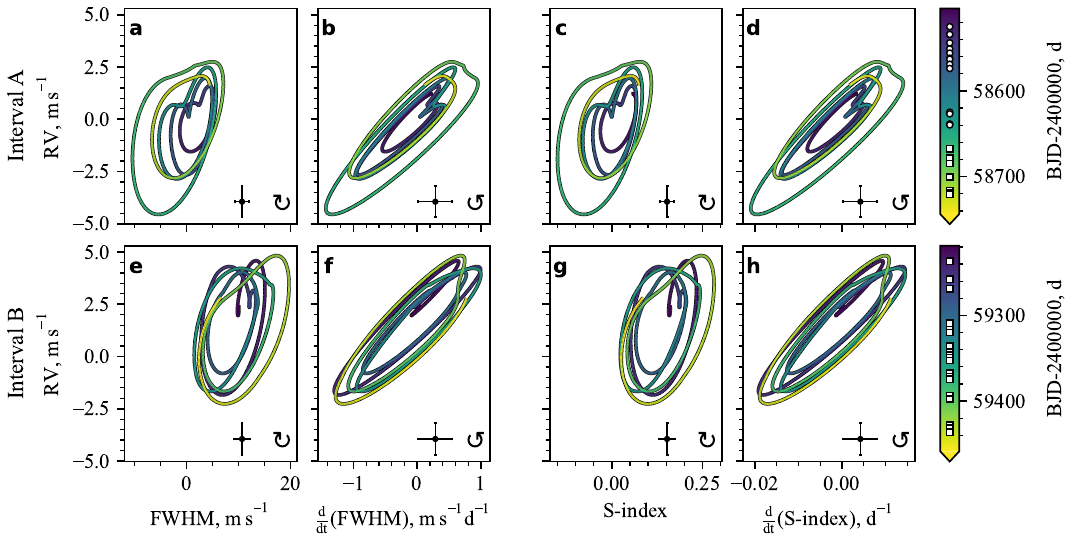}}
    \caption{
    Phase-space trajectories built from our best-model predictions in Fig.~\ref{fig:bestmodel_fit}, in the same selected intervals. We compare RV with:
    (a,b) FWHM,
    (c,d) the time derivative of FWHM,
    (e,f) S-index,
    (g,h) the time derivative of S-index.
    Time runs from blue to yellow, as defined by the colorbars at the right-hand side. The same colorbars contain the temporal location of ESPRESSO18 (circles) and ESPRESSO19 (squares) measurements. Every trajectory comes with a mark at the bottom right that guides the direction of evolution, as well as the median model uncertainty in each dimension.
    }
    \label{fig:bestmodel_trajectory}
\end{figure*}

The two highlighted intervals in Fig.~\ref{fig:bestmodel_fit}: intervals A and B, are among the most densely measured in our dataset. Each spans \SI{240}{\day}, i.e. about $5P_\text{rot}$; and they are also \SI{715}{\day} apart, i.e. about $0.4P_\text{cyc}$ apart. The dense ESPRESSO sampling allows us to peek into the stellar activity within a few stellar-rotation periods -- and the large gap between intervals offers the opportunity to compare them head-to-head and to look for long-term changes. We took our best-model predictions from Fig.~\ref{fig:bestmodel_fit}, and computed the time derivatives of activity indicators, again at \SI{0.1}{\day} sampling.\footnote{We computed activity-indicator time derivatives numerically through the centred finite-difference method \mbox{$f'(x)|_{x_i}=[f(x_{i+1})-f(x_{i-1})]/2h$}, with \mbox{$h=\SI{0.1}{\day}$}. We used the matrix representation of this method to transform the conditional mean vector and the covariance matrix of each activity indicator so as to get to the conditional mean vector and the covariance matrix of its time derivative.}
Then, we plotted the predicted RVs against both indicators and their time derivatives, as a function of time, in Fig.~\ref{fig:bestmodel_trajectory}. For instance, Fig.~\ref{fig:bestmodel_trajectory}a takes the predicted RV against time (Fig.~\ref{fig:bestmodel_fit}a) as well as the predicted FWHM against time (Fig.~\ref{fig:bestmodel_fit}e), and plots them against one another, with time being colour-coded. These plots visualise the phase spaces between RV and activity indicators (or the time derivatives of the latter), and reveal the phase-space trajectories that GJ~526 follows over time.

Figure~\ref{fig:bestmodel_trajectory} shows phase-space trajectories over approximately $5P_\text{rot}$. At the same time, all trajectories comprise five closed contours that somewhat repeat themselves. This means that for each geometrical revolution of GJ~526 around itself (one $P_\text{rot}$), we have one revolution in phase space (one contour). But there is more -- the trajectories reveal intricate features in stellar activity, which we will now briefly address. By virtue of the apparent (and model-assumed) positive correlation between FWHM and S-index, we restrict ourselves to a discussion between RV, FWHM, and the time derivative of FWHM.

The first phase-space trajectory (Fig.~\ref{fig:bestmodel_trajectory}a; RV and FWHM in Interval~A) already reveals interesting features of the stellar activity of GJ~526. Here, the trajectory revolves clockwise, but contours do not maintain the same size. In fact, the trajectory starts out small (Fig.~\ref{fig:bestmodel_trajectory}a; dark-blue contour), then unwinds to a contour of a phase-space area that is several times larger at around \SIrange{2458620}{2458660}{\bjd}, and then winds inwards again in the end of the interval. This behaviour can be verified by inspection of the peak-to-peak distance evolution in Fig.~\ref{fig:bestmodel_fit}a and Fig.~\ref{fig:bestmodel_fit}e, but it only becomes clear in a phase-space trajectory plot. It is worth mentioning that not all contours are convex -- the first three contours all have an inward curl (Fig.~\ref{fig:bestmodel_trajectory}a, near the origin). This is a manifestation of the `flip-flop' effect,
i.e. a cyclic shift of active regions to the other parts of the stellar surface (\citealp{Berdyugina1999}; see also \citealp{Olah2006,Hackman2013}), which can be again verified in the RV- and FWHM time series (Figs.~\ref{fig:bestmodel_fit}a,e).

In Interval B, almost half a cycle later, the RV-FWHM trajectory continues to revolve clockwise (Fig.~\ref{fig:bestmodel_trajectory}e). We catch a remnant of a flip-flop effect in the very beginning of the interval, as well as a second curl near \SI{2459270}{\bjd}. Here, the whole phase-space trajectory is translated upwards and rightwards relative to Interval A. This suggests a positive shift in RV and FWHM, which we attribute to the long-term cycle itself. In this interval, we do not observe any apparent growth or shrinkage between contours, and their apparent centres seem to move at a slower rate than in Interval A.

We observe similar trajectory structures in the phase space formed by RV and the first FWHM derivative (Fig.~\ref{fig:bestmodel_trajectory}b,f). Here, contours are less circular, and have a more positive covariance. The trajectories revolve anticlockwise -- opposite to the RV-FWHM ones. Curls can again be sighted, but this time they are easily spotted only in Interval A (Fig.~\ref{fig:bestmodel_trajectory}b). In the same interval, we observe a similar situation of unwinding to a larger contour around \SIrange{2458620}{2458660}{\bjd}, which winds inwards again in the end of the interval. This is consistent with Fig.~\ref{fig:bestmodel_trajectory}a, since a large variance of the FWHM derivative would imply a large variance in FWHM as well. Finally, the apparent centre of contours only shifts upwards from Interval A to Interval B. This suggests that while FWHM appears to be offset significantly within a cycle (Figs.~\ref{fig:bestmodel_trajectory}a,e), the derivative does not.

The phase-space trajectories of GJ~526 are a natural argument against the use of a direct correlation analysis between RV and activity indicators. Depending on the temporal sampling of an observing campaign, measurements may `land' on different parts of the trajectory -- meaning that an observer may get a positive, a negative, or no correlation at all.
The need to factor in the times of measurement as well as measurements themselves has been noted in earlier works (e.g. \citealp{CollierCameron2019}). Phase-space trajectories satisfy this need, and may open new doors for analysis.

We report the following median model uncertainties in Fig.~\ref{fig:bestmodel_trajectory} in RV, FWHM, FWHM derivative, S-index, and S-index derivative:
\SI{0.74}{\metre\per\second},
\SI{1.38}{\metre\per\second},
\SI{0.27}{\metre\per\second\per\day},
\num{0.0202}, and
 \SI{0.0039}{\per\day} in Interval A; as well as
\SI{0.76}{\metre\per\second},
\SI{1.45}{\metre\per\second},
\SI{0.26}{\metre\per\second\per\day},
\num{0.0212}, and
 \SI{0.0039}{\per\day} in Interval B (marked with errorbars in subplots).
Although these median uncertainties are larger than the intricate curls we observe in the phase spaces, we note that most curls are significant compared to their local uncertainties (e.g. near
\SI{2458550}{\bjd},
\SI{2458640}{\bjd}, and
\SI{2459270}{\bjd}; Fig.~\ref{fig:bestmodel_fit}a,b). We anticipate that the precision of a phase-space trajectory at a given time $t$ grows with the number of measurements within one timescale $\tau$ from $t$, as well as with the precision of measurements themselves. As such, trajectory analyses should be conducted in well-measured intervals, even if only in one physical quantity, provided that the GP hyperparameters are sufficiently well-defined.

We used the same model to analyse data from several instruments, and these data may in principle come with their own chromatic effects. In our case, however, the wavelength ranges are similar enough so that no significant chromaticity is expected for an early M dwarf (\SIrange{380}{788}{\nano\metre} for ESPRESSO, \SIrange{380}{690}{\nano\metre} for HARPS, \SIrange{300}{630}{\nano\metre} for HIRES RVs). There is only one timeseries overlap -- that of HIRES and HARPS in RV and S-index -- and we do not observe a strong discrepancy in behaviour or scatter in neither case (Fig.~\ref{fig:timeseries}a,e). The FWHM may have a different mean value depending on the instrumental profile, but relative variations have been shown to behave consistently while following astronomical phenomena. This consistency has been recently demonstrated between ESPRESSO and HARPS, in both RV and FWHM (\citealp{SuarezMascareno2025}, Fig.~F.1; \citealp{Hobson2025}, Fig.~6). The S-index is defined as a flux ratio of regions with strict wavelength boundaries (Eq. \ref{eq:s_index}), and is consequently expected to be invariant.

\subsection{Detection limits for additional companions}

Works related to exoplanetary searches often attempt to assess their sensitivity to any undiscovered planets in a system. This gives rise to different tests which probe the so-called `detection limit' of a certain study. However, the very design of these tests changes the meaning of a detection limit. For example, one way to assess sensitivity is to inject a random circular-planet signal in the raw RV data, and then fit the best model again onto the modified raw time series, using the median posterior had already been inferred in the main analysis. The best model may absorb the injected signal, or may leave it out, causing a strong peak in the residual GLSP in the second case. One can therefore inject different sines, one at a time, and determine sensitivity from some aggregate statistics of the injected signal after fitting (e.g. \citealp{Konacki2009,Howard2016}).
Such tests are referred to as `injection-recovery' tests, and their detection limits are fundamentally based on the inability of their models to leave an injected RV signal out.

Other tests seek for an extra planet directly in data, with no injection nor recovery. For example, \citet{Faria2020} used the sampled posteriors of their models with planets, and propagated them into a period-mass distribution. We used a variant of this test. We started with our best model, augmented it to include an extra circular-planet signal, and sampled for its parameters, having fixed all other model parameters except the RV zero offset $\alpha_0$ and the RV jitters $J_{i,0}$.\footnote{The uneven sampling of a potential planet introduces its own RV offset, hence the need of sampling for $\alpha_0$. Such planet may also change the model jitters in RV, hence the need of sampling for $J_{i,0}$.}
We did this for many bins of the orbital-period space, with a separate inference run per bin. Then, for each bin, we considered the $\pm3\sigma$ percentiles of the RV semi-amplitude posterior of the extra planet. The $+3\sigma$ percentile can be interpreted as the maximum semi-amplitude an undiscovered planet would have with a $3\sigma$ probability, if it existed at all -- and we took it as our detection limit. The $-3\sigma$ percentile tells of any significant signals that stand well away from the zero, which could deserve attention in further studies.

\begin{figure}
    \centering
    \resizebox{\hsize}{!}{\includegraphics{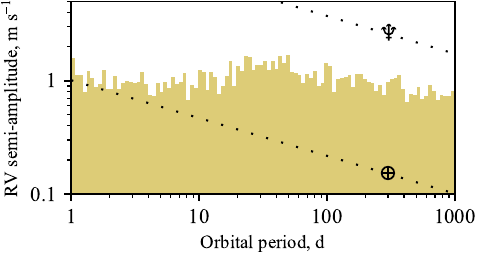}}
    \caption{
    Detection limits of GJ~526. Shaded bands show the $3\sigma$ RV semi-amplitude posterior of a modelled planet in our best model, in 100 log-spaced orbital-period bins. Two dotted lines show the RV semi-amplitudes that an Earth-mass planet and an Neptune-mass planet would inject in the system.
    }
    \label{fig:model_detectability}
\end{figure}

We took the orbital-period interval \mbox{$\mathcal{U}_{\log}(1,1000)\,\unit\day$}, and partitioned it in 100 bins. In every of these simpler `bin-models', we used an RV semi-amplitude of
\mbox{$\mathcal{U}(0,5)\,\unit{\metre\per\second}$}, the same RV-offset prior as in our model grid, and a reduced RV-jitter prior of \mbox{$\mathcal{U}(0,5)\,\unit{\metre\per\second}$} for all instruments. To speed up inference, in every instance of $N_\text{param}$ bin-model parameters, we required $20N_\text{param}$ live points (instead of $40N_\text{param}$ in the model grid), and integrated until 10\% of the log-likelihood integral was left in the iterative remainder (instead of the default 1\% in \textsc{Ultranest}). Figure~\ref{fig:model_detectability} displays the $\pm 3\sigma$ percentiles of our 100 bins in the parameter space.
 Overall, our $+3\sigma$ detection limit stands around \SI{1}{\metre\per\second}, consistent with our model jitters (\SIrange{0.88}{2.26}{\metre\per\second}). This limit shows a very modest degradation to \SI{1.5}{\metre\per\second} at periods in the range \SIrange{20}{80}{\day}, on the same scale as $P_\text{rot}$. With the data at hand, we are already sensitive to the detection of any Neptune-mass planets up to \SI{e3}{\day}, as they induce larger semi-amplitudes than the $3\sigma$ detection limit. On the other hand, the $-3\sigma$ percentile never exceeds \SI{10}{\centi\metre\per\second}.

Our detection-limit test takes place in the space formed by the RV semi-amplitude and the orbital period. We took the derived stellar parameters and their uncertainties in Table~\ref{tab:stellar_parameters}, and used them to transform the semi-amplitude to minimum mass, as well as the orbital period to semi-major axis (SMA). Under such transformations, we can reject the presence of planets up to a certain mass for a given SMA. Our test disfavours the presence of planets
beyond \SI{5}{\earthmass} in the interval \SIrange{0.02}{0.16}{\au}, and
beyond \SI{8}{\earthmass} in the interval \SIrange{0.16}{1.4}{\au}.
At the same time, the vicinity of GJ~526 to the Sun allows the proper-motion anomaly technique to place similar limits to longer-period planets: \citet{Kervella2022} found that there appear to be no planets beyond \SIrange{25}{30}{\earthmass} in the interval \SIrange{3}{10}{\au}. Such combinations of techniques allow to screen for planets over wide ranges of orbital separations.

\subsection{False-inclusion probability tests}

We assessed the significance of any RV signals that escaped our modelling following \citet{Hara2022}, hereafter \citetalias{Hara2022}. Their formalism uses the posterior distribution of an inference run to compute the probability of having no planets within a given element in the orbital-frequency space. \citetalias{Hara2022} refers to this metric as the false inclusion probability (FIP). We took the posteriors of our MEP model with one circular-orbit planet (Fig.~\ref{fig:model_grid}; bottom left) and invoked the FIP formalism between \SIrange{1}{1000}{\day} for an angular-frequency step \mbox{$\Delta\omega=2\pi/5T_\text{bl}$}, where $T_\text{bl}$ is our dataset baseline. We found several inconspicuous peaks in the range \SIrange{10}{100}{\day}, with the most significant being assigned a FIP of about 70\% at a period near $P_\text{rot}$. No signals passed the 1\% threshold suggested by \citetalias{Hara2022} that supports a planetary detection.

\section{Conclusions}\label{sec:conclusions}
We perform a stellar-activity analysis of GJ~526 in three physical quantities (RV, FWHM, S-index) on a rich and multi-faceted velocimetry that the HIRES, HARPS, and ESPRESSO instruments provide. Through simultaneously fitting a multi-dimensional GP model in all three quantities, we constrain the stellar-rotation period of GJ~526 to \mbox{$P_\text{rot}=48.7\pm 0.3\,\unit\day$} and deduce that its active regions are dark spots on the stellar surface. We demonstrate that 65 ESPRESSO measurements alone replicate these results with an acceptable precision. The combined HIRES, HARPS and ESPRESSO spectroscopy suggests that GJ~526 exhibits weak magnetic activity with a period of $P_\text{cyc}=1680^{+50}_{-40}\,\unit\day$. Current evidence does not support the existence of a planetary companion around GJ~526. We estimate that if such companion existed, it would likely have a RV semi-amplitude smaller than \SI{1.5}{\metre\per\second}.

We propose the use of phase-space trajectories as a way to illustrate and understand stellar activity (Sect.~\ref{sec:rv_activity_trajectories}). We argue that their use allows the identification of less apparent details in activity. Unlike correlations between RV and activity indicators, phase-space trajectories contain temporal information and can be informative as long as the region of interest contains enough points to guarantee a precise model prediction. As an example, we qualitatively compare phase-space trajectories in RV, FWHM, and S-index in two intervals separated by almost half a magnetic cycle. Further work is required to determine the usefulness of such phase-space trajectories. For instance, it may be the case that the direction of revolution, or the geometrical features of phase-space trajectories (e.g. angular speed or area) carry some physical meaning. These prospects can be explored through a trajectory comparison with observational data of other cool dwarfs, or with synthetic data from stellar-activity simulation suites.

\begin{acknowledgements}
AKS acknowledges the support of a fellowship from the ”la Caixa” Foundation (ID 100010434). The fellowship code is LCF/BQ/DI23/11990071.
AKS, JIGH, ASM, NN and RR acknowledge financial support from the Spanish Ministry of Science and Innovation (MICINN) project PID2020-117493GB-I00 and from the Government of the Canary Islands project ProID2020010129.
NN acknowledges funding from Light Bridges for the Doc-
toral Thesis ``Habitable Earth-like planets with ESPRESSO and
NIRPS'', in cooperation with the Instituto de Astrofísica de Canarias, and the use of Indefeasible Computer Rights (ICR) be-
ing commissioned at the ASTRO POC project in the Island of
Tenerife, Canary Islands (Spain). The ICR-ASTRONOMY used
for his research was provided by Light Bridges in cooperation
with Hewlett Packard Enterprise (HPE).
Co-funded by the European Union (ERC, FIERCE, 101052347). Views and opinions expressed are however those of the author(s) only and do not necessarily reflect those of the European Union or the European Research Council. Neither the European Union nor the granting authority can be held responsible for them. This work was supported by FCT - Fundação para a Ciência e a Tecnologia through national funds by these grants:
UIDB/04434/2020 DOI: 10.54499/UIDB/04434/2020,
UIDP/04434/2020 DOI: 10.54499/UIDP/04434/2020,
PTDC/FIS-AST/4862/2020, UID/04434/2025,
2022.04048.PTDC DOI: 10.54499/2022.04048.PTDC.
CJAPM also acknowledges FCT and POCH/FSE (EC) support through Investigador FCT Contract 2021.01214.CEECIND/CP1658/CT0001 (DOI 10.54499/2021.01214.CEECIND/CP1658/CT0001).
AC-G is funded by the Spanish Ministry of Science through MCIN/AEI/10.13039/501100011033 grant PID2019-107061GB-C61.

This work includes data collected by the TESS mission and made use of \textsc{lightkurve}, a Python package for Kepler and TESS data analysis \citep{lightkurve}. Funding for the TESS mission is provided by the NASA's Science Mission Directorate.
This research has made use of NASA's Astrophysics Data System Bibliographic Services. We made use of NASA’s Astrophysics Data System, which is operated by the California Institute of Technology, under contract with the National Aeronautics and Space Administration under the Exoplanet Exploration Program.
This research has made use of the \textsc{SIMBAD} \citep{simbad} and \textsc{VizieR} \citep{vizier} databases, both operated at CDS, Strasbourg, France.
Additionally, we made use of the Revised Version of the New Catalogue of Suspected Variable Stars \citep{nsv2}.

We thank the anonymous referee for their time and effort. AKS acknowledges Jean-Baptiste Delisle for the prompt support on some technical queries regarding \textsc{s+leaf}.
We used the following \textsc{Python} packages for data analysis and
visualisation:
\textsc{Astropy} \citep{astropy},
\textsc{Matplotlib} \citep{matplotlib},
\textsc{nieva} (Stefanov et al., in prep.),
\textsc{NumPy} \citep{numpy},
\textsc{pandas} \citep{pandas1,pandas2},
\textsc{SciPy} \citep{scipy},
\textsc{seaborn} \citep{seaborn},
\textsc{s+leaf} \citep{spleaf1,spleaf2} and
\textsc{ultranest} \citep{ultranest}.
This manuscript was written and compiled in \textsc{Overleaf}. All presented analysis was conducted on \textsc{Ubuntu} machines. The bulk of modelling and inference was done on
the Diva cluster (192 Xeon E7-4850 \SI{2.1}{\giga\hertz} CPUs; \SI{4.4}{\tera\byte} RAM) at Instituto de Astrofísica de Canarias, Tenerife, Spain. \end{acknowledgements}

\bibliographystyle{aa} \bibliography{zotero_bibtex} 

\let\cleardoublepage\clearpage
\begin{appendix}
\section{GJ~526 photometry}\label{sec:photometry}
\subsection{ASAS-SN photometry}\label{sec:photometry_asassn}
\begin{figure}
    \centering
    \resizebox{\hsize}{!}{\includegraphics{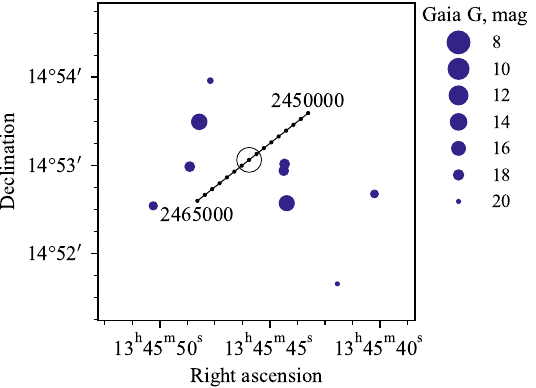}}
    \caption{
        The proper motion of GJ~526 (black hollow circle) on the sky plane relative to other stars in the vicinity (blue circles). The line shows the trajectory of GJ~526 in the interval \SIrange{2450000}{2465000}{\julianday}, with small circular markers at every \SI{1000}{\julianday}. Stars are size-coded relative to their brightness in Gaia G \citep{gaiaDR3}.
    }
    \label{fig:gaia_propermotion}
\end{figure}

We used the ASAS-SN Sky Patrol online service\footnote{\url{https://asas-sn.osu.edu/}} to inspect the photometry of GJ~526 at hand. Our preliminary check of the field revealed that there are no significant contaminators that can bias photometry (Fig.~\ref{fig:gaia_propermotion}). However, the total proper motion of GJ~526 approaches \SI{2.3}{\arcsec\per\year} \citepalias{gaiaDR3}, which is comparable to the pixel size of the detector (\SI{8}{\arcsec}). Past works address this issue through a computation of many separate light curves associated with smaller temporal intervals (e.g. \citealp{Trifonov2021,Castro-Gonzalez2023,Damasso2023}). We went forward with the same procedure. We did split the region \SIrange{2456000}{2461000}{\julianday} in timestamps of \SI{250}{\julianday}. Then, for each time stamp, we computed the expected position of GJ~526 and used the Sky Patrol service to compute an individual light curve for the temporal region defined by $\pm\SI{125}{\julianday}$ around the time stamp. We used the saturated-stars photometry method as described in \citet{Winecki2024}. Finally, we concatenated individual light curves together, removed all points of lunar separation below \SI{90}{\deg} and clipped the remainder at $3\sigma$. This procedure yielded 354 data points in Sloan g, and 99 data points in Johnson V.

\begin{figure*}
    \centering
    \resizebox{\hsize}{!}{\includegraphics{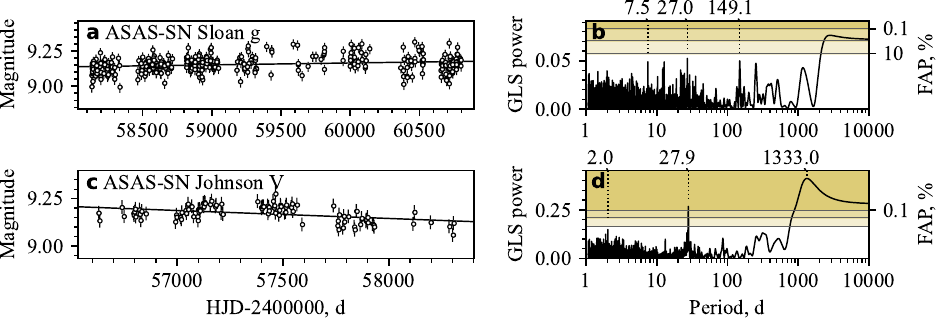}}
    \caption{
    ASAS-SN photometry of GJ~526 that we discuss in this work, in: (a) Sloan g, (c) Johnson V. (b,d) The associated Lomb-Scargle periodograms of data after linear detrending.
    }
    \label{fig:asassn_timeseries}
\end{figure*}

Fig.~\ref{fig:asassn_timeseries} displays our ASAS-SN data. The rms of measurements against the median photometric uncertainty is \SI{55}{\milli\mag} against \SI{37}{\milli\mag} for Sloan g; and \SI{41}{\milli\mag} against \SI{37}{\milli\mag} for Johnson V. The observed rms can explain the variations measured by \citet{Shakhovskaya1971} \mbox{($\Delta m_V=\SI{40}{\milli\mag}$)}. Both timeseries follow an apparent trend, which manifested in the GLSPs as strong powers at infinite period. We did a linear fit on photometry (as highlighted by lines in Figs.~\ref{fig:asas_timeseries}a,c), which partly, but not completely suppressed these infinite-period powers. Sloan g data appears featureless, except a large power in the range \SIrange{2000}{3000}{\day}. Johnson V data, on the other hand, contains two significant peaks: one at \SI{27.9}{\day} and another at \SI{1333}{\day}. The former is close to the lunar orbital period, but we note a strong WF peak at the same location in these timeseries,\footnote{In fact, this WF peak is injected from our own pre-processing step that masks by lunar separation.}
meaning that the \SI{27.9}{\day} signature in photometry likely comes from the interaction between the WF peak in question at the infinite-period power, as described in \eqref{eq:alias_equation}. After higher-order polynomial fitting, the \SI{27.9}{\day} peak disappeared, which confirmed our suspicions. The \SI{1333}{\day} peak is somewhat close to our measurements of a magnetic cycle at period \mbox{$P_\text{cyc}=1670\pm 40\,\unit\day$}.

\subsection{ASAS photometry}

\begin{figure*}
    \centering
    \resizebox{\hsize}{!}{\includegraphics{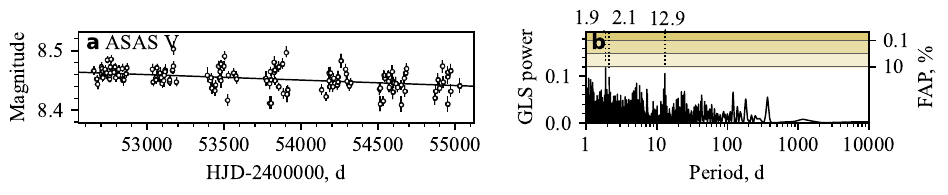}}
    \caption{
    ASAS-SN photometry of GJ~526 that we discuss in this work.
    (a) ASAS V timeseries.
    (b) The associated Lomb-Scargle periodogram of data after linear detrending.
    }
    \label{fig:asas_timeseries}
\end{figure*}

We took all GJ~526 ASAS photometry of quality flag `A' or `B', following \citet{SuarezMascareno2016a}. Additionally, we required a minimum lunar separation of \SI{90}{\deg}, much like our ASAS-SN photometry. This query returned 150 data points in ASAS V with an rms of \SI{17}{\milli\mag} and a median uncertainty of \SI{8}{\milli\mag}. Photometry appears relatively stable and follows an inconspicuous linear trend with time (Fig.~\ref{fig:asas_timeseries}a). After subtraction of said trend, no strong signatures remain in the GLSP (Fig.~\ref{fig:asas_timeseries}b).

\subsection{TESS photometry}

\begin{figure*}
    \centering
    \resizebox{\hsize}{!}{\includegraphics{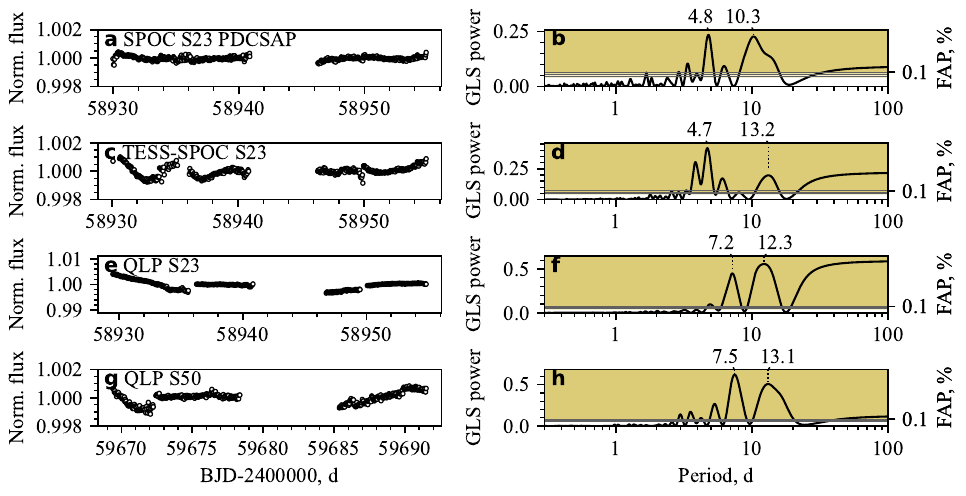}}
    \caption{
    TESS photometry of GJ~526 that we discuss in this work, from the following pipelines:
    (a) SPOC \mbox{PDCSAP} from Sector 23,
    (c) TESS-SPOC from Sector 23,
    (e) QLP from Sector 23,
    (g) QLP from Sector 50.
    (b,d,f,h) The associated Lomb-Scargle periodograms of data.
    }
    \label{fig:tess_timeseries}
\end{figure*}

There is available GJ~526 photometry in two TESS sectors: S23 and S50. S23 photometry was obtained in the primary mission and was analysed by the official pipeline, SPOC \citep{spoc} -- as well as by TESS-SPOC \citep{tess_spoc} and QLP \citep{tess_qlp1,tess_qlp2,tess_qlp3,tess_qlp4} on the 30-minute FFIs. As a result, there are light-curves of \SI{2}{\minute} and \SI{30}{\minute} cadence in this sector alone. S50 photometry took place during the extended TESS mission, and was reduced by the QLP pipeline only, producing a light-curve of \SI{30}{\minute} cadence. All of this amounts to four data products in total from both sectors. We present their photometry in Fig.~\ref{fig:tess_timeseries}.

We took all aforementioned data, clipped it at $3\sigma$ and normalised it through division by the flux mean of each timeseries. We show the normalises timeseries Fig.~\ref{fig:tess_timeseries}. Overall, we observe long-term trends in both S23 and S50, which are qualitatively similar and follow the same order: (i) a linear decline; (ii) a short curl upwards; (iii) a plateau until the first half of photometric coverage; (iv) a smooth rise in the second half. S23 and S50 are separated by more than \SI{700}{\day}, which makes any physical explanation unfavourable. Among all GLSPs, we note strong signals near \SI{5}{\day}, \SI{7}{\day} and \SI{13}{\day}. All of these signatures fall at periods where light-curve modulations can be attributed to instrument systematics \citep{Fetherolf2023}.

\section{Supplementary material}
\begin{figure*}
    \centering
    \begin{minipage}{\columnwidth}
    \resizebox{\hsize}{!}{\includegraphics{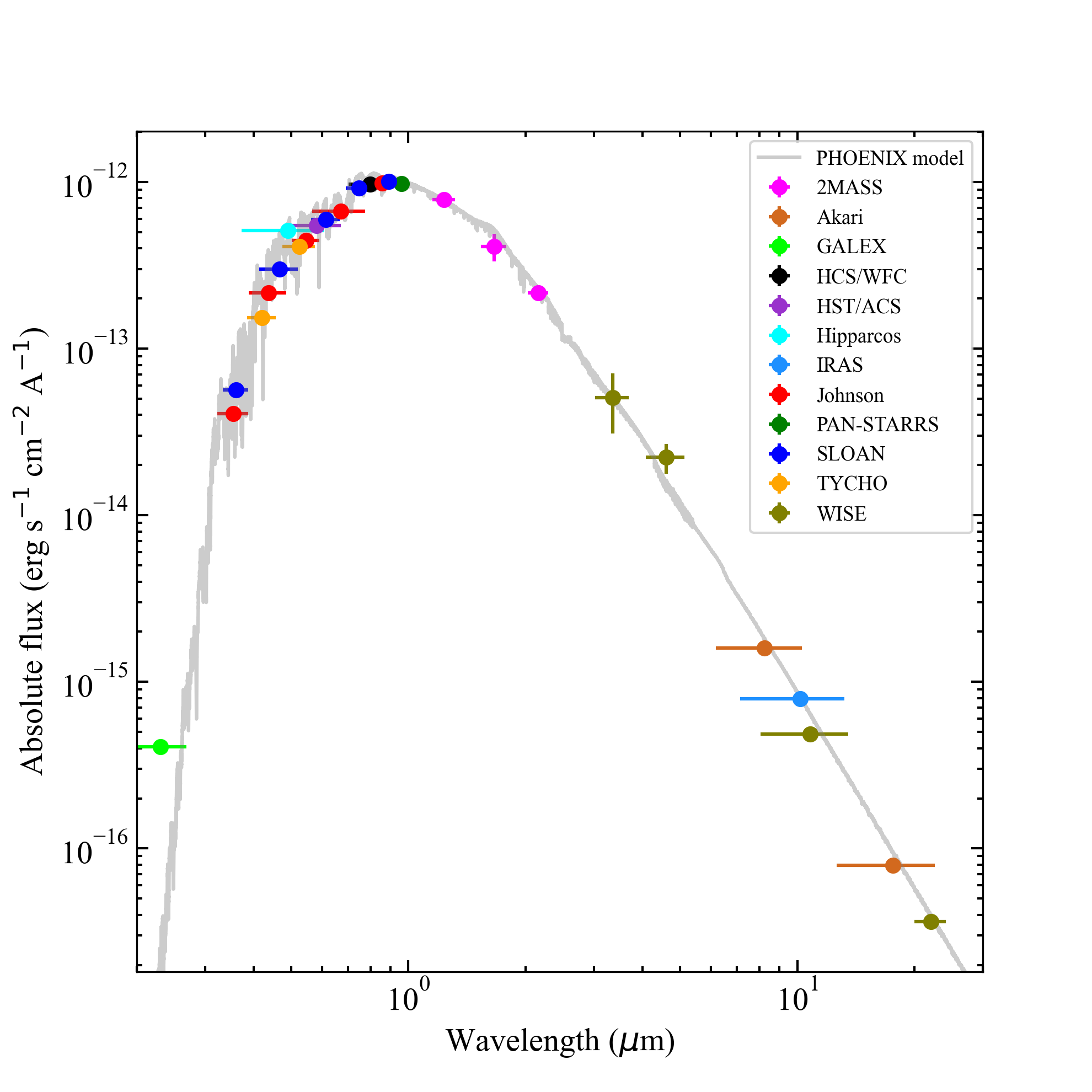}}
    \caption{
    Observed spectral energy distribution of GJ~526 (coloured dots) from the ultraviolet through the mid-infrared. We compare those against a solar-metallicity, $\log g = 5.0$, \SI{3700}{\kelvin} PHOENIX model (gray line).
    }\label{fig:mariarosa}
    \end{minipage}
    \hfill
    \vspace{3em}
    \resizebox{\hsize}{!}{\includegraphics{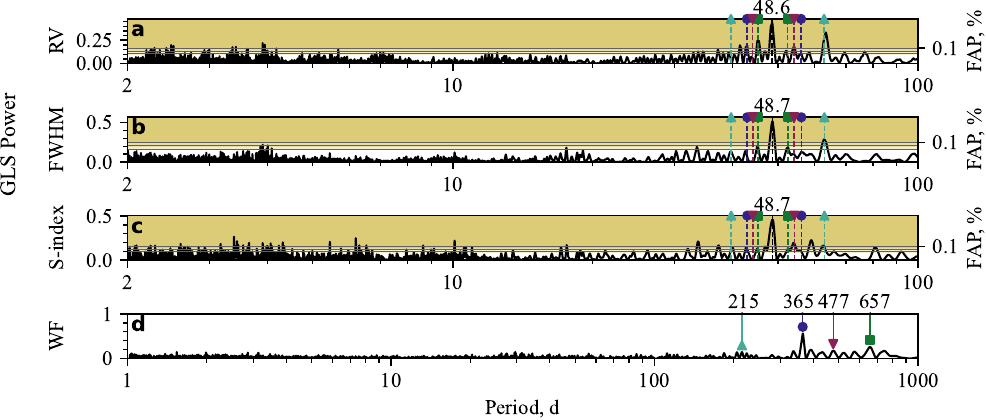}}
    \caption{
    Alias analysis of raw-timeseries: (a) RV, (b) FWHM, (c) S-index. We focus on the aliases of the \SI{48.6}{\day} signal.
    (d)~The window function reveals four prominent peaks at:
    \SI{215}{\day} (upward teal triangle),
    \SI{365}{\day} (blue circle),
    \SI{475}{\day} (downward red triangle), and
    \SI{635}{\day} (green square).
    Relevant aliases in data periodograms are plotted with dashed lines of a corresponding colour and marker.
    }\label{fig:alias_analysis}
\end{figure*}
\clearpage
\begin{figure*}
    \centering
    \resizebox{\hsize}{!}{\includegraphics{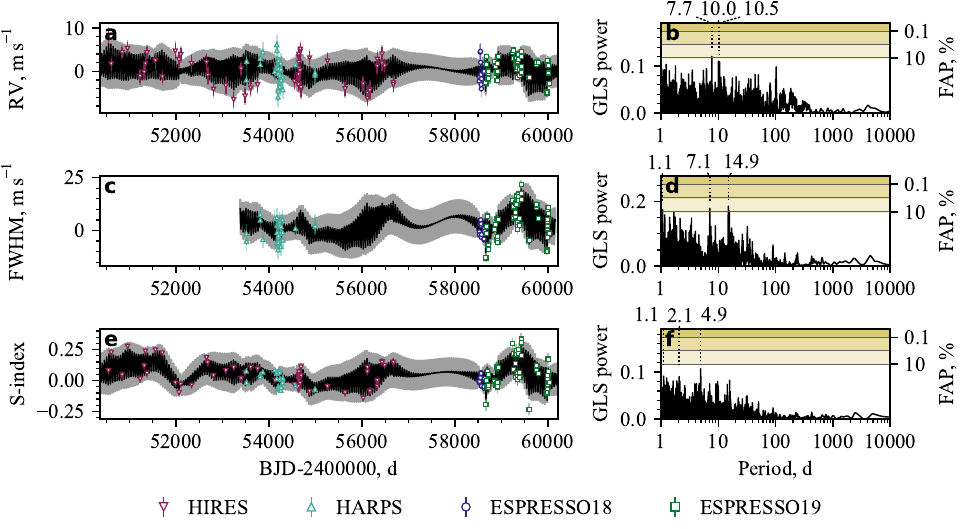}}
    \caption{
    Raw time series (markers) against our best model (black; LTF+GP), over the whole dataset baseline in:
    (a) RV,
    (c) FWHM,
    (e) S-index.
    Data points come with two error bars: one assigned by the instrument (saturated), and another that includes the model jitter (semi-transparent). 
    (b,d,f) Associated GLSPs of the residual time series of our best model, accounting for the model jitter.
    }
    \label{fig:bestmodel_fit_full}
\end{figure*}
 \begin{figure*}
    \centering
    \resizebox{\hsize}{!}{\includegraphics{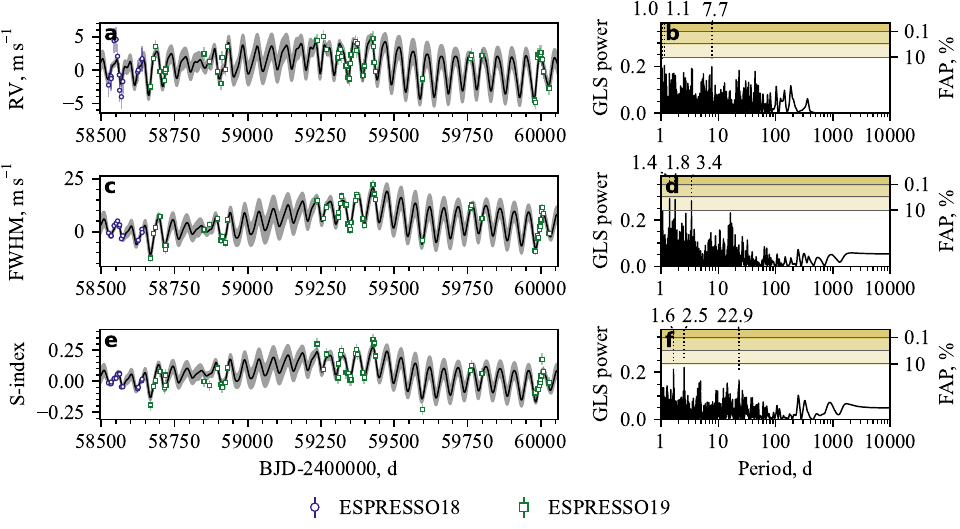}}
    \caption{
    Equivalent to Fig.~\ref{fig:bestmodel_fit_full}, but for a derivative of the best model that was fit on ESPRESSO data alone.
    }
    \label{fig:esprmodel_fit_full}
\end{figure*}
 \clearpage

\begin{figure*}
    \centering
    \resizebox{\hsize}{!}{\includegraphics{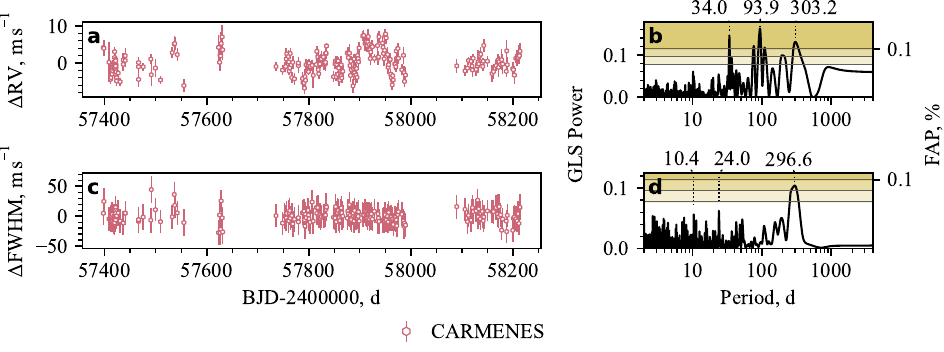}}
    \caption{
    CARMENES $3\sigma$ clipped timeseries of mean-subtracted:
    (a) RV,
    (c) FWHM. (b,d) Associated wide-period GLSPs of RV and FWHM. Three FAP levels: 10\%, 1\% and 0.1\%, split GLSP ordinates in bands of different colour. We highlight the three most prominent peaks in each GLSP.
    }
    \label{fig:carm_timeseries}
\end{figure*}
 \begin{figure*}
    \centering
    \resizebox{\hsize}{!}{\includegraphics{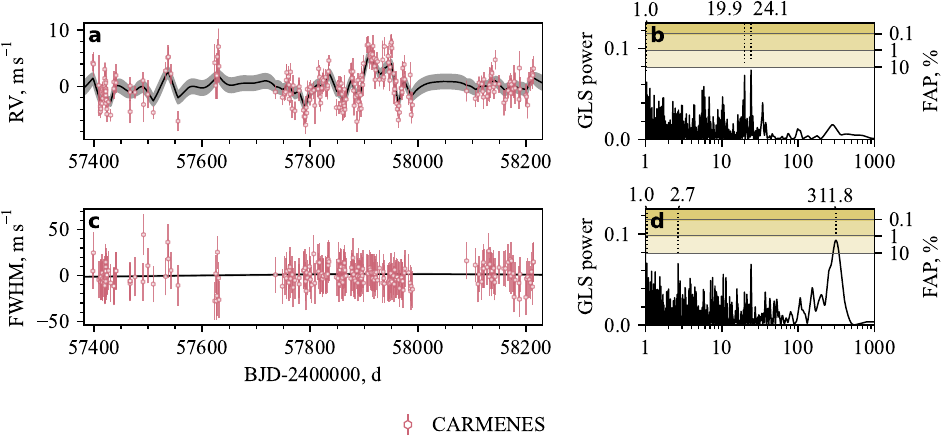}}
    \caption{
    Equivalent to Fig.~\ref{fig:bestmodel_fit_full}, but for a derivative of the best model that was fit on CARMENES $3\sigma$ clipped data alone.
    }
    \label{fig:carmmodel_fit_full}
\end{figure*}
 \clearpage

\begin{table*}
    \centering
    \caption{
        Parameter posteriors of our best model on all data, against the same set up on ESPRESSO data alone.
    }
    \label{tab:bestmodel_posterior}
    
    \begin{tabular}{lcclcc}
    \hline \hline
    Parameter name &
    Symbol &
    Unit &
    Prior &
    \multicolumn{2}{c}{Posterior} \\ \cline{5-6}
     &
     &
     &
     &
    \textbf{All-data} &
    ESPRESSO \\ \hline
    \textbf{LTF parameters} \\
    Period &
    $P_\text{cyc}$ &
    \unit\day &
    $\mathcal{U}\left(1000, 2000\right)$ &
    $1680^{+50}_{-40}$ &
    $1720^{+190}_{-210}$ \\
    RV phase &
    $\varphi_\text{cyc, 0}$ &
    - &
    $\mathcal{U}\left(0, 1\right)$\tablefootmark{w} &
    $0.843^{+0.118}_{-0.129}$ &
    $0.872^{+0.149}_{-0.177}$ \\
    RV semi-amplitude &
    $k_\text{cyc, 0}$ &
    \unit{\metre\per\second} &
    $\mathcal{U}\left(0, 10\right)$ &
    $0.51^{+0.35}_{-0.33}$ &
    $0.54^{+0.47}_{-0.37}$ \\
    FWHM phase &
    $\varphi_\text{cyc, 1}$ &
    - &
    $\mathcal{U}\left(0, 1\right)$\tablefootmark{w} &
    $0.429^{+0.085}_{-0.089}$ &
    $0.448^{+0.145}_{-0.132}$ \\
    FWHM semi-amplitude &
    $k_\text{cyc, 1}$ &
    \unit{\metre\per\second} &
    $\mathcal{U}\left(0, 10\right)$ &
    $2.73^{+1.51}_{-1.32}$  &
    $2.34^{+1.68}_{-1.44}$  \\
    S-index phase &
    $\varphi_\text{cyc, 2}$ &
    - &
    $\mathcal{U}\left(0, 1\right)$\tablefootmark{w} &
    $0.548^{+0.252}_{-0.248}$ &
    $0.854^{+0.342}_{-0.268}$ \\
    S-index semi-amplitude &
    $k_\text{cyc, 2}$ &
    - &
    $\mathcal{U}\left(0, 0.1\right)$ &
    $0.019^{+0.019}_{-0.013}$  &
    $0.023^{+0.025}_{-0.016}$  \\
    RV zero-order correction &
    $\alpha_{0}$ &
    \unit{\metre\per\second} &
    $\mathcal{N}(\mu_\text{lm},200\sigma_\text{lm})$ &
    $0.47^{+0.85}_{-0.83}$ &
    $0.60^{+1.25}_{-1.27}$ \\
    FWHM zero-order correction &
    $\alpha_{1}$ &
    \unit{\metre\per\second} &
    $\mathcal{N}(\mu_\text{lm},200\sigma_\text{lm})$ &
    $3.89^{+3.28}_{-3.23}$ &
    $4.11^{+5.05}_{-5.25}$ \\
    S-index zero-order correction &
    $\alpha_{2}$ &
    - &
    $\mathcal{N}(\mu_\text{lm},200\sigma_\text{lm})$ &
    $0.050\pm 0.048$ &
    $0.051^{+0.075}_{-0.077}$ \\
    \textbf{Dataset parameters} \\
    ESPRESSO19 RV offset &
    $O_{1,0}$ &
    \unit{\metre\per\second} &
    $\mathcal{N}(0,5\sigma_0)$ &
    $-1.06\pm 0.74$ &
    $-1.17^{+0.93}_{-0.94}$ \\
    HIRES RV offset &
    $O_{2,0}$ &
    \unit{\metre\per\second} &
    $\mathcal{N}(0,5\sigma_0)$ &
    $0.00^{+1.00}_{-1.01}$ &
    - \\
    HARPS RV offset &
    $O_{3,0}$ &
    \unit{\metre\per\second} &
    $\mathcal{N}(0,5\sigma_0)$ &
    $-0.05^{+1.13}_{-1.12}$ &
    - \\
    ESPRESSO19 FWHM offset &
    $O_{1,1}$ &
    \unit{\metre\per\second} &
    $\mathcal{N}(0,5\sigma_1)$ &
    $-4.63^{+2.30}_{-2.21}$ &
    $-5.04^{+3.02}_{-3.20}$ \\
    HARPS FWHM offset &
    $O_{3,1}$ &
    \unit{\metre\per\second} &
    $\mathcal{N}(0,5\sigma_1)$ &
    $-0.19^{+4.30}_{-4.10}$ &
    - \\
    ESPRESSO19 S-index offset &
    $O_{1,2}$ &
    - &
    $\mathcal{N}(0,5\sigma_2)$ &
    $\left(-6.22^{+3.44}_{-3.38}\right)\times 10^{-2}$ &
    $\left(-6.57^{+4.55}_{-4.84}\right)\times 10^{-2}$ \\
    HIRES S-index offset &
    $O_{2,2}$ &
    - &
    $\mathcal{N}(0,5\sigma_2)$ &
    $\left(-4.63^{+5.39}_{-5.44}\right)\times 10^{-2}$ &
    - \\
    HARPS S-index offset &
    $O_{3,2}$ &
    - &
    $\mathcal{N}(0,5\sigma_2)$ &
    $\left(0.07^{+6.09}_{-5.77}\right)\times 10^{-2}$ &
    - \\
    ESPRESSO18 RV jitter &
    $J_{0,0}$ &
    \unit{\metre\per\second} &
    $\exp[\mathcal{N}(\ln\sigma_0,\ln\sigma_0)]$ &
    $1.82^{+0.47}_{-0.35}$ &
    $1.82^{+0.48}_{-0.36}$ \\
    ESPRESSO19 RV jitter &
    $J_{1,0}$ &
    \unit{\metre\per\second} &
    $\exp[\mathcal{N}(\ln\sigma_0,\ln\sigma_0)]$ &
    $0.85^{+0.21}_{-0.22}$ &
    $1.01^{+0.20}_{-0.23}$ \\
    HIRES RV jitter &
    $J_{2,0}$ &
    \unit{\metre\per\second} &
    $\exp[\mathcal{N}(\ln\sigma_0,\ln\sigma_0)]$ &
    $1.20^{+0.71}_{-0.65}$ &
    - \\
    HARPS RV jitter &
    $J_{3,0}$ &
    \unit{\metre\per\second} &
    $\exp[\mathcal{N}(\ln\sigma_0,\ln\sigma_0)]$ &
    $2.21^{+0.43}_{-0.36}$ &
    - \\
    ESPRESSO18 FWHM jitter &
    $J_{0,1}$ &
    \unit{\metre\per\second} &
    $\mathcal{U}_{\log}\left(10^{-3}, 10^{3}\right)$ &
    $1.02^{+0.34}_{-0.27}$ &
    $1.00^{+0.34}_{-0.26}$ \\
    ESPRESSO19 FWHM jitter &
    $J_{1,1}$ &
    \unit{\metre\per\second} &
    $\mathcal{U}_{\log}\left(10^{-3}, 10^{3}\right)$ &
    $1.82^{+0.34}_{-0.29}$ &
    $1.78^{+0.35}_{-0.31}$ \\
    HARPS FWHM jitter &
    $J_{3,1}$ &
    \unit{\metre\per\second} &
    $\mathcal{U}_{\log}\left(10^{-3}, 10^{3}\right)$ &
    $4.25^{+0.77}_{-0.62}$ &
    - \\
    ESPRESSO18 S-index jitter &
    $J_{0,2}$ &
    - &
    $\mathcal{U}_{\log}\left(10^{-3}, 10^{3}\right)$ &
    $\left(1.27^{+0.55}_{-0.45}\right)\times 10^{-2}$ &
    $\left(1.46^{+0.58}_{-0.51}\right)\times 10^{-2}$ \\
    ESPRESSO19 S-index jitter &
    $J_{1,2}$ &
    - &
    $\mathcal{U}_{\log}\left(10^{-3}, 10^{3}\right)$ &
    $\left(4.77^{+0.62}_{-0.54}\right)\times 10^{-2}$ &
    $\left(4.73^{+0.66}_{-0.65}\right)\times 10^{-2}$ \\
    HIRES S-index jitter &
    $J_{2,2}$ &
    - &
    $\mathcal{U}_{\log}\left(10^{-3}, 10^{3}\right)$ &
    $\left(3.14^{+0.75}_{-0.59}\right)\times 10^{-2}$ &
    - \\
    HARPS S-index jitter &
    $J_{3,2}$ &
    - &
    $\mathcal{U}_{\log}\left(10^{-3}, 10^{3}\right)$ &
    $\left(0.43^{+0.64}_{-0.27}\right)\times 10^{-2}$ &
    - \\
    \textbf{Stellar-activity hyperparameters} \\
    Timescale &
    $\tau$ &
    \unit\day &
    $\mathcal{U}_{\log}\left(20, 10^{4}\right)$ &
    $277^{+109}_{-82}$ &
    $227^{+281}_{-109}$ \\
    Period &
    $P_\text{rot}$ &
    \unit\day &
    $\mathcal{U}(40.1,64.1)$ &
    $48.7\pm 0.3$ &
    $48.5^{+0.6}_{-1.0}$ \\
    Sinescale (harmonic complexity) &
    $\eta$ &
    - &
    $\mathcal{U}_{\log}\left(10^{-2}, 10^{2}\right)$ &
    $0.53^{+0.13}_{-0.10}$ &
    $0.94^{+0.78}_{-0.32}$ \\
    RV amplitude &
    $A_0$ &
    \unit{\metre\per\second} &
    $\mathcal{U}(-10^3,10^3)$ &
    $1.45^{+0.30}_{-0.26}$ &
    $2.21^{+2.05}_{-0.69}$ \\
    RV gradient amplitude &
    $B_0$ &
    \unit{\metre\per\second\per\day} &
    $\mathcal{U}(-10^3,10^3)$ &
    $19.9^{+4.0}_{-3.2}$ &
    $23.1^{+23.3}_{-7.3}$ \\
    FWHM amplitude &
    $A_1$ &
    \unit{\metre\per\second} &
    $\mathcal{U}_{\log}\left(10^{-3}, 10^{3}\right)$ &
    $7.26^{+1.13}_{-0.86}$ &
    $9.76^{+9.10}_{-2.78}$ \\
    S-index amplitude &
    $A_2$ &
    \unit{\metre\per\second} &
    $\mathcal{U}_{\log}\left(10^{-3}, 10^{3}\right)$ &
    $0.106^{+0.015}_{-0.012}$ &
    $0.133^{+0.137}_{-0.042}$ \\
    \hline
    \end{tabular}
    \tablefoot{
    \tablefoottext{w}{Wrapped parameter.}
    Reported uncertainties reflect the 16\textsuperscript{th} and the 84\textsuperscript{th} percentiles. Offsets are defined relative to ESPRESSO18. The standard deviation of measurements in physical quantity $j$ is denoted $\sigma_j$.
    }
\end{table*} \begin{sidewaystable*}
    \small
    \centering
    \caption{
    Posterior comparison of our best model on all data (MEP; Table~\ref{tab:bestmodel_posterior}) against planet-free models in our grid that used an ESP kernel with 2, 3, or 4 harmonics.
    }
    \label{tab:bestmodel_posterior_comparison}
    
    \begin{tabular}{lcclcccc}
    \hline \hline
    Parameter name &
    Symbol &
    Unit &
    Prior &
    \multicolumn{4}{c}{Posterior} \\ \cline{5-8}
    &&&&
    \textbf{MEP} &
    ESP2 &
    ESP3 &
    ESP4 \\ \hline
    \textbf{LTF parameters} \\
    Period &
    $P_\text{cyc}$ &
    \unit\day &
    $\mathcal{U}\left(1000, 2000\right)$ &
    $1680^{+50}_{-40}$ &
    $1680\pm 50$ &
    $1680\pm 60$ &
    $1680^{+60}_{-50}$ \\
    RV phase &
    $\varphi_\text{cyc, 0}$ &
    - &
    $\mathcal{U}\left(0, 1\right)$\tablefootmark{w} &
    $0.843^{+0.118}_{-0.129}$ &
    $0.854^{+0.123}_{-0.131}$ &
    $0.848^{+0.156}_{-0.175}$ &
    $0.861^{+0.140}_{-0.156}$ \\
    RV semi-amplitude &
    $k_\text{cyc, 0}$ &
    \unit{\metre\per\second} &
    $\mathcal{U}\left(0, 10\right)$ &
    $0.51^{+0.35}_{-0.33}$ &
    $0.50^{+0.36}_{-0.33}$ &
    $0.38^{+0.34}_{-0.26}$ &
    $0.41^{+0.36}_{-0.28}$ \\
    FWHM phase &
    $\varphi_\text{cyc, 1}$ &
    - &
    $\mathcal{U}\left(0, 1\right)$\tablefootmark{w} &
    $0.429^{+0.085}_{-0.089}$ &
    $0.434^{+0.087}_{-0.085}$ &
    $0.425\pm 0.101$ &
    $0.427^{+0.100}_{-0.093}$ \\
    FWHM semi-amplitude &
    $k_\text{cyc, 1}$ &
    \unit{\metre\per\second} &
    $\mathcal{U}\left(0, 10\right)$ &
    $2.73^{+1.51}_{-1.32}$  &
    $2.54^{+1.27}_{-1.25}$  &
    $2.23^{+1.25}_{-1.24}$  &
    $2.33^{+1.26}_{-1.24}$  \\
    S-index phase &
    $\varphi_\text{cyc, 2}$ &
    - &
    $\mathcal{U}\left(0, 1\right)$\tablefootmark{w} &
    $0.548^{+0.252}_{-0.248}$ &
    $0.577^{+0.265}_{-0.280}$ &
    $0.664^{+0.272}_{-0.287}$ &
    $0.681^{+0.276}_{-0.255}$ \\
    S-index semi-amplitude &
    $k_\text{cyc, 2}$ &
    - &
    $\mathcal{U}\left(0, 0.1\right)$ &
    $0.019^{+0.019}_{-0.013}$  &
    $0.015^{+0.016}_{-0.011}$  &
    $0.015^{+0.016}_{-0.010}$  &
    $0.015^{+0.016}_{-0.011}$  \\
    RV zero-order correction &
    $\alpha_{0}$ &
    \unit{\metre\per\second} &
    $\mathcal{N}(\mu_\text{lm},200\sigma_\text{lm})$ &
    $0.47^{+0.85}_{-0.83}$ &
    $0.57^{+0.86}_{-0.84}$ &
    $0.45\pm 0.82$ &
    $0.43\pm 0.82$ \\
    FWHM zero-order correction &
    $\alpha_{1}$ &
    \unit{\metre\per\second} &
    $\mathcal{N}(\mu_\text{lm},200\sigma_\text{lm})$ &
    $3.89^{+3.28}_{-3.23}$ &
    $4.28^{+3.19}_{-3.12}$ &
    $3.95^{+2.98}_{-3.00}$ &
    $3.96^{+2.89}_{-2.86}$ \\
    S-index zero-order correction &
    $\alpha_{2}$ &
    - &
    $\mathcal{N}(\mu_\text{lm},200\sigma_\text{lm})$ &
    $0.050\pm 0.048$ &
    $0.056\pm 0.048$ &
    $0.053\pm 0.044$ &
    $0.052^{+0.043}_{-0.042}$ \\
    \textbf{Dataset parameters} \\
    ESPRESSO19 RV offset &
    $O_{1,0}$ &
    \unit{\metre\per\second} &
    $\mathcal{N}(0,5\sigma_0)$ &
    $-1.06\pm 0.74$ &
    $-1.17^{+0.87}_{-0.86}$ &
    $-1.02^{+0.82}_{-0.83}$ &
    $-0.95^{+0.82}_{-0.84}$ \\
    HIRES RV offset &
    $O_{2,0}$ &
    \unit{\metre\per\second} &
    $\mathcal{N}(0,5\sigma_0)$ &
    $0.00^{+1.00}_{-1.01}$ &
    $-0.30^{+1.01}_{-1.02}$ &
    $-0.23^{+0.96}_{-0.97}$ &
    $-0.19\pm 0.96$ \\
    HARPS RV offset &
    $O_{3,0}$ &
    \unit{\metre\per\second} &
    $\mathcal{N}(0,5\sigma_0)$ &
    $-0.05^{+1.13}_{-1.12}$ &
    $-0.10^{+1.13}_{-1.09}$ &
    $0.07^{+1.07}_{-1.04}$ &
    $0.09^{+1.06}_{-1.05}$ \\
    ESPRESSO19 FWHM offset &
    $O_{1,1}$ &
    \unit{\metre\per\second} &
    $\mathcal{N}(0,5\sigma_1)$ &
    $-4.63^{+2.30}_{-2.21}$ &
    $-4.97^{+3.14}_{-2.95}$ &
    $-4.57^{+2.72}_{-2.63}$ &
    $-4.50\pm 2.57$ \\
    HARPS FWHM offset &
    $O_{3,1}$ &
    \unit{\metre\per\second} &
    $\mathcal{N}(0,5\sigma_1)$ &
    $-0.19^{+4.30}_{-4.10}$ &
    $-1.13^{+3.87}_{-3.79}$ &
    $-0.64^{+3.78}_{-3.77}$ &
    $-0.63^{+3.76}_{-3.65}$ \\
    ESPRESSO19 S-index offset &
    $O_{1,2}$ &
    - &
    $\mathcal{N}(0,5\sigma_2)$ &
    $\left(-6.22^{+3.44}_{-3.38}\right)\times 10^{-2}$ &
    $\left(-6.69^{+4.77}_{-4.49}\right)\times 10^{-2}$ &
    $\left(-6.26^{+4.05}_{-3.88}\right)\times 10^{-2}$ &
    $\left(-6.07^{+3.85}_{-3.90}\right)\times 10^{-2}$ \\
    HIRES S-index offset &
    $O_{2,2}$ &
    - &
    $\mathcal{N}(0,5\sigma_2)$ &
    $\left(-4.63^{+5.39}_{-5.44}\right)\times 10^{-2}$ &
    $\left(-5.14^{+5.15}_{-5.16}\right)\times 10^{-2}$ &
    $\left(-4.74^{+4.76}_{-4.84}\right)\times 10^{-2}$ &
    $\left(-4.74^{+4.67}_{-4.68}\right)\times 10^{-2}$ \\
    HARPS S-index offset &
    $O_{3,2}$ &
    - &
    $\mathcal{N}(0,5\sigma_2)$ &
    $\left(0.07^{+6.09}_{-5.77}\right)\times 10^{-2}$ &
    $\left(-1.63^{+5.54}_{-5.55}\right)\times 10^{-2}$ &
    $\left(-1.09^{+5.31}_{-5.35}\right)\times 10^{-2}$ &
    $\left(-1.03^{+5.14}_{-5.21}\right)\times 10^{-2}$ \\
    ESPRESSO18 RV jitter &
    $J_{0,0}$ &
    \unit{\metre\per\second} &
    $\exp[\mathcal{N}(\ln\sigma_0,\ln\sigma_0)]$ &
    $1.82^{+0.47}_{-0.35}$ &
    $1.82^{+0.48}_{-0.37}$ &
    $1.90^{+0.47}_{-0.35}$ &
    $1.94^{+0.48}_{-0.36}$ \\
    ESPRESSO19 RV jitter &
    $J_{1,0}$ &
    \unit{\metre\per\second} &
    $\exp[\mathcal{N}(\ln\sigma_0,\ln\sigma_0)]$ &
    $0.85^{+0.21}_{-0.22}$ &
    $0.92\pm 0.20$ &
    $0.85^{+0.20}_{-0.22}$ &
    $0.71\pm 0.23$ \\
    HIRES RV jitter &
    $J_{2,0}$ &
    \unit{\metre\per\second} &
    $\exp[\mathcal{N}(\ln\sigma_0,\ln\sigma_0)]$ &
    $1.20^{+0.71}_{-0.65}$ &
    $1.48^{+0.80}_{-0.84}$ &
    $1.63^{+0.65}_{-0.83}$ &
    $1.58^{+0.64}_{-0.78}$ \\
    HARPS RV jitter &
    $J_{3,0}$ &
    \unit{\metre\per\second} &
    $\exp[\mathcal{N}(\ln\sigma_0,\ln\sigma_0)]$ &
    $2.21^{+0.43}_{-0.36}$ &
    $2.32^{+0.45}_{-0.40}$ &
    $2.24^{+0.47}_{-0.42}$ &
    $2.19^{+0.48}_{-0.44}$ \\
    ESPRESSO18 FWHM jitter &
    $J_{0,1}$ &
    \unit{\metre\per\second} &
    $\mathcal{U}_{\log}\left(10^{-3}, 10^{3}\right)$ &
    $1.02^{+0.34}_{-0.27}$ &
    $1.05^{+0.34}_{-0.25}$ &
    $1.09^{+0.35}_{-0.28}$ &
    $1.09^{+0.35}_{-0.29}$ \\
    ESPRESSO19 FWHM jitter &
    $J_{1,1}$ &
    \unit{\metre\per\second} &
    $\mathcal{U}_{\log}\left(10^{-3}, 10^{3}\right)$ &
    $1.82^{+0.34}_{-0.29}$ &
    $1.95^{+0.38}_{-0.31}$ &
    $1.75^{+0.40}_{-0.33}$ &
    $1.78^{+0.39}_{-0.33}$ \\
    HARPS FWHM jitter &
    $J_{3,1}$ &
    \unit{\metre\per\second} &
    $\mathcal{U}_{\log}\left(10^{-3}, 10^{3}\right)$ &
    $4.25^{+0.77}_{-0.62}$ &
    $4.24^{+0.76}_{-0.61}$ &
    $4.29^{+0.78}_{-0.62}$ &
    $4.29^{+0.79}_{-0.62}$ \\
    ESPRESSO18 S-index jitter &
    $J_{0,2}$ &
    - &
    $\mathcal{U}_{\log}\left(10^{-3}, 10^{3}\right)$ &
    $\left(1.27^{+0.55}_{-0.45}\right)\times 10^{-2}$ &
    $\left(1.54^{+0.58}_{-0.47}\right)\times 10^{-2}$ &
    $\left(0.87^{+0.55}_{-0.32}\right)\times 10^{-2}$ &
    $\left(0.84^{+0.53}_{-0.31}\right)\times 10^{-2}$ \\
    ESPRESSO19 S-index jitter &
    $J_{1,2}$ &
    - &
    $\mathcal{U}_{\log}\left(10^{-3}, 10^{3}\right)$ &
    $\left(4.77^{+0.62}_{-0.54}\right)\times 10^{-2}$ &
    $\left(4.62^{+0.64}_{-0.55}\right)\times 10^{-2}$ &
    $\left(4.70^{+0.63}_{-0.55}\right)\times 10^{-2}$ &
    $\left(4.76^{+0.64}_{-0.56}\right)\times 10^{-2}$ \\
    HIRES S-index jitter &
    $J_{2,2}$ &
    - &
    $\mathcal{U}_{\log}\left(10^{-3}, 10^{3}\right)$ &
    $\left(3.14^{+0.75}_{-0.59}\right)\times 10^{-2}$ &
    $\left(3.20^{+0.83}_{-0.62}\right)\times 10^{-2}$ &
    $\left(3.24^{+0.85}_{-0.62}\right)\times 10^{-2}$ &
    $\left(3.26^{+0.84}_{-0.63}\right)\times 10^{-2}$ \\
    HARPS S-index jitter &
    $J_{3,2}$ &
    - &
    $\mathcal{U}_{\log}\left(10^{-3}, 10^{3}\right)$ &
    $\left(0.43^{+0.64}_{-0.27}\right)\times 10^{-2}$ &
    $\left(0.43^{+0.64}_{-0.28}\right)\times 10^{-2}$ &
    $\left(0.40^{+0.65}_{-0.25}\right)\times 10^{-2}$ &
    $\left(0.42^{+0.69}_{-0.26}\right)\times 10^{-2}$ \\
    \textbf{Stellar-activity hyperparameters} \\
    Timescale &
    $\tau$ &
    \unit\day &
    $\mathcal{U}_{\log}\left(20, 10^{4}\right)$ &
    $277^{+109}_{-82}$ &
    $68^{+22}_{-19}$ &
    $79^{+23}_{-20}$ &
    $81^{+22}_{-18}$ \\
    Period &
    $P_\text{rot}$ &
    \unit\day &
    $\mathcal{U}(40.1,64.1)$ &
    $48.7\pm 0.3$ &
    $49.2^{+1.7}_{-1.5}$ &
    $49.2^{+1.2}_{-0.9}$ &
    $49.3^{+1.1}_{-0.9}$ \\
    Sinescale (harmonic complexity) &
    $\eta$ &
    - &
    $\mathcal{U}_{\log}\left(10^{-2}, 10^{2}\right)$ &
    $0.53^{+0.13}_{-0.10}$ &
    $0.60^{+0.16}_{-0.14}$ &
    $0.58\pm 0.08$ &
    $0.55^{+0.08}_{-0.07}$ \\
    RV amplitude &
    $A_0$ &
    \unit{\metre\per\second} &
    $\mathcal{U}(-10^3,10^3)$ &
    $1.45^{+0.30}_{-0.26}$ &
    $1.42^{+0.28}_{-0.25}$ &
    $1.33^{+0.26}_{-0.23}$ &
    $1.33^{+0.25}_{-0.22}$ \\
    RV gradient amplitude &
    $B_0$ &
    \unit{\metre\per\second\per\day} &
    $\mathcal{U}(-10^3,10^3)$ &
    $19.9^{+4.0}_{-3.2}$ &
    $19.6^{+3.9}_{-3.3}$ &
    $17.8^{+3.6}_{-3.0}$ &
    $17.3^{+3.4}_{-2.8}$ \\
    FWHM amplitude &
    $A_1$ &
    \unit{\metre\per\second} &
    $\mathcal{U}_{\log}\left(10^{-3}, 10^{3}\right)$ &
    $7.26^{+1.13}_{-0.86}$ &
    $6.72^{+0.87}_{-0.73}$ &
    $6.78^{+0.83}_{-0.71}$ &
    $6.73^{+0.82}_{-0.71}$ \\
    S-index amplitude &
    $A_2$ &
    \unit{\metre\per\second} &
    $\mathcal{U}_{\log}\left(10^{-3}, 10^{3}\right)$ &
    $0.106^{+0.015}_{-0.012}$ &
    $0.101^{+0.012}_{-0.010}$ &
    $0.100^{+0.011}_{-0.010}$ &
    $0.099^{+0.011}_{-0.010}$ \\ \hline
    \end{tabular}
    \tablefoot{
    \tablefoottext{w}{Wrapped parameter.}
    Reported uncertainties reflect the 16\textsuperscript{th} and the 84\textsuperscript{th} percentiles. Offsets are defined relative to ESPRESSO18. The standard deviation of measurements in physical quantity $j$ is denoted $\sigma_j$.
    }
\end{sidewaystable*}  \end{appendix}
\end{document}